\documentclass[10pt]{article}

\usepackage{amsmath}
\usepackage{amssymb}

\usepackage{graphicx}

\usepackage{cite}
\usepackage{hyperref}
\usepackage{color}

\usepackage{setspace}
\usepackage{colortbl}

\usepackage[labelfont=bf,labelsep=period,justification=raggedright]{caption}

\makeatletter
\renewcommand{\@biblabel}[1]{\quad#1.}
\makeatother
\def\R2Lurl#1#2{\mbox{\href{#1}{\tt #2}}}

\usepackage[section] {placeins}

\usepackage{soul}

\date{}

\begin{document}

\begin{flushleft}
{\Large
\textbf{Dynamics of conflicts in Wikipedia}
}
\\
Taha Yasseri$^{1\ast}$,
Robert Sumi$^{1}$,
Andr\'as Rung$^{1}$,
Andr\'as Kornai$^{1,2}$,
J\'{a}nos Kert\'{e}sz$^{1}$
\\
\bf{1} Department of Theoretical Physics, Budapest University of Technology and Economics, Budapest, Hungary.
\\
\bf{2} Computer and Automation Research Institute, Hungarian Academy of Sciences, Budapest, Hungary.
\\
$\ast$ E-mail: yasseri@phy.bme.hu
\end{flushleft}


\section*{Abstract}
In this work we study the dynamical features of editorial wars in Wikipedia (WP).
Based on our previously established algorithm, we build up samples of controversial and peaceful
articles and analyze the temporal characteristics of the activity in these samples. On short time
scales, we show that there is a clear correspondence between conflict and burstiness of activity
patterns, and that memory effects play an important role in controversies. On long time scales,
we identify three distinct developmental patterns for the overall behavior of the articles.
We are able to distinguish cases eventually leading to consensus from those cases where a
compromise is far from achievable. Finally, we analyze discussion networks and conclude that
edit wars are mainly fought by few editors only.

\section*{Introduction}\label{sec:intro}

New media such as the internet and the web enable entirely new ways of
collaboration, opening unprecedented opportunities for handling tasks of
extraordinary size and complexity. Such collaborative schemes have already been
used to solve challenges in software engineering \cite{gnu} and mathematics
\cite{gowers2009}. Understanding the laws of internet-based collaborative
value production is of great importance.

Perhaps the most prominent example of such value production is Wikipedia (WP),
a free, collaborative, multilingual internet encyclopedia \cite{wiki-main}. WP
evolves without the supervision of a pre-selected expert team, its voluntary
editors define the rules and maintain the quality. WP has grown beyond other
encyclopedias both in size and in use, having unquestionably become the number
one reference in practice. Although criticism has been continuously expressed
concerning its reliability and accuracy, partly because the editorial policy
is in favor of consensus over credentials \cite{wiki-policy}, independent
studies have shown that, as early as in 2005, science articles in WP and
Encyclopedia Britannica were of comparable quality \cite{giles2005}.  As every
edit and discussion post is saved and available, WP is particularly well
suited to study internet-based collaborative processes. Indeed, WP has been
studied extensively from different aspects including the growth of content and
community \cite{voss2005,ortega2007}, coverage \cite{halavais2008,kittur2009}
and evolution of the hyperlink networks
\cite{buriol2006,capocci2006,zlatic2006,zlatic2011,ratkiewicz2010a}, the
extraction of semantic networks \cite{strube2006,ponzetto2007,medelyan2009},
linguistic studies \cite{tyers2008,sharoff2008,yasseri2012c}, user reputation
\cite{javanmardi2010b} and collaboration quality
\cite{javanmardi2010a,kimmons2011}, vandalism detection
\cite{potthast2008,smets2008,adler2011}, and the social aspects of the
editor community
\cite{hu2007,leskovec2010,mcdonald2011,laniado2011a,massa2011,laniado2011b}.

Usually, different editors constructively extend each other's text, correct
minor errors and mistakes until a consensual article emerges -- this is the
most natural, and by far the most common, way for a WP entry to be
developed \cite{wilkinson2007}. Good examples include (WP articles will be cited in {\tt typewriter font}
throughout the text)
{\tt Benjamin Franklin, Pumpkin} or {\tt
  Helium}. As we shall see, in the English WP close to 99\% of the articles
result from this rather smooth, constructive process. However, the
development of WP articles is not always peaceful and collaborative,  there
are sometimes heavy fights called {\it edit wars} between groups representing
opposing opinions. Schneider et al. \cite{schneider2010}
estimated that in the English WP, among the highly edited or highly viewed
articles (these notions are strongly correlated, see \cite{ratkiewicz2010}),
about 12\% of discussions are devoted to reverts and
vandalism, suggesting that the WP development process for articles of major
interest is highly contentious. The WP community has created a full system of
measures to resolve conflict situations, including the so called ``three
revert rule'' (see {\tt Wikipedia:Edit warring}), locking articles for
non-registered editors, tagging controversial articles, and temporal or final
banning of malevolent editors. It is against this rich backdrop of explicit
rules, explicit or implicit regulations, and unwritten conventions that the
present paper undertakes to investigate a fundamental part of the
collaborative value production, how conflicts emerge and get resolved.

The first order of business is to construct an automated procedure to identify
controversial articles. For a human reader the simplest way to do so is to go
to the discussion (talk) pages of the articles, which often show the typical
signatures of conflicts as known from social psychology \cite{samson2010}. The
length of the discussion page could already be considered a good indicator of
conflict: the more severe the conflict, the longer the talk page is expected
to be (this will be shown in detail later).  However, this feature is very
language dependent: while conflicts are indeed fought out in detail on
discussion pages in the English WP, German editors do not use this vehicle for
the same purpose.  Moreover, there are WPs, e.g. the Hungarian one, where
discussion pages are always rather sparse, rarely mentioning the actual
arguments. Clearly the discussion page alone is not an appropriate source to
identify conflicts if we aim at a general, multi-lingual, culture-independent
indicator.

Conflicts in WP were studied previously both on the article and on the user
level. Kittur et al. \cite{bongwon2007,kittur2007} and Vuong et
al. \cite{vuong2008} measured controversiality by counting the
``controversial'' tag in the history of an article, and compared other
possible metrics to that. It should be noted, however, that this is at best a
one-sided measure as highly disputed pages such as {\tt Gdansk} or {\tt
  Euthanasia} in the English WP lack such tags, and the situation is even
worse in other WPs. In \cite{kittur2007}, different page metrics like the
number of reverts, the number of revisions etc. were compared to the tag
counts and in \cite{vuong2008} the number of deleted words between users were
counted and a ``Mutual Reinforcement Principle'' \cite{zha2002} was used to
measure how controversial a given article is.  Clearly, there are several
features of an article which correlate with its controversiality, making it
highly non-trivial to choose an appropriate indicator.  Some papers try to
detect the negative ``conflict'' links between WP editors in a given article
and, based on this, attempt to classify editors into groups. The main idea of
the method used by Kittur et al. \cite{kittur2007} is to relate the severity
of the conflict between two editors to the number of reverts they carry out on
each other's versions.  In a more recent study \cite{brandes2008,brandes2009},
Brandes et al. counted the number of deleted words between editors and used
this as a measure of controversy.

There is no question that reverting a part of an article expresses strong
disagreement, but sometimes this is just related to eliminating vandalized
texts, while in other cases it is related to conflict about the contents of
the article. Here we are interested in the second case and it will be one of
our goals to distinguish between deeper conflict and mere vandalism. Beyond
identifying conflict pages and edit wars, we aim at relating different
properties of the articles to their level of controversiality. In the Methods
section we describe the dataset, summarize our conflict identification method,
and relate it to other measures proposed in the literature. In the main body
of the paper we analyze the temporal evolution of conflicts both on the micro and
the macro timescales and, based on that, we try to categorize them.

\section*{Methods}\label{sec:methods}

To analyze edit wars in WP first we need to be able to detect the articles
where significant debates occur.  For the human viewer of page histories it
is evident that an article such as {\tt Liancourt Rocks}, discussing a group
of small islets claimed by both Korea and Japan, or the article on {\tt
  Homosexuality} were the subject of major edit wars. Yet articles with a
similar number or relative proportion of edits such as {\tt Benjamin Franklin}
or {\tt Pumpkin} were, equally evidently to the human reader, developed
peacefully.  For our conflict detection method (previously reported in
\cite{sumi2011a}, \cite{sumi2011b}), similar to most pattern recognition tasks
such as speech or character recognition, we take human judgment to be the gold
standard or ``truth'' against which machine performance is to be judged.
How human judgment is solicited is discussed in Text~S1.

The whole structured dataset and the implementation of the ranking algorithm described below, along with the raw results,
are available at {\it WikiWarMonitor} webpage: \url{http://wwm.phy.bme.hu/}.

\subsection*{Dataset}
Our analysis is based on the January 2010 dump of the English WP \cite{wiki-dumps}, 
which contains
all the versions of all pages up to that date. The dataset originally contains
3.2 M articles, but we have filtered out all short (less than 1,000
characters) and evidently conflict-free (less than 100 edits) articles,
leaving a final set of around 223 k articles.

\subsection*{Detecting edit wars}

Our detection method is entirely based on statistical features of edits and is
therefore independent of language characteristics.  This makes possible both
inter-cultural comparisons and cross-language checks and validation.

\subsubsection*{Revert maps}

\begin{figure}[ht]
\begin{center}
\includegraphics[width=0.75\textwidth]{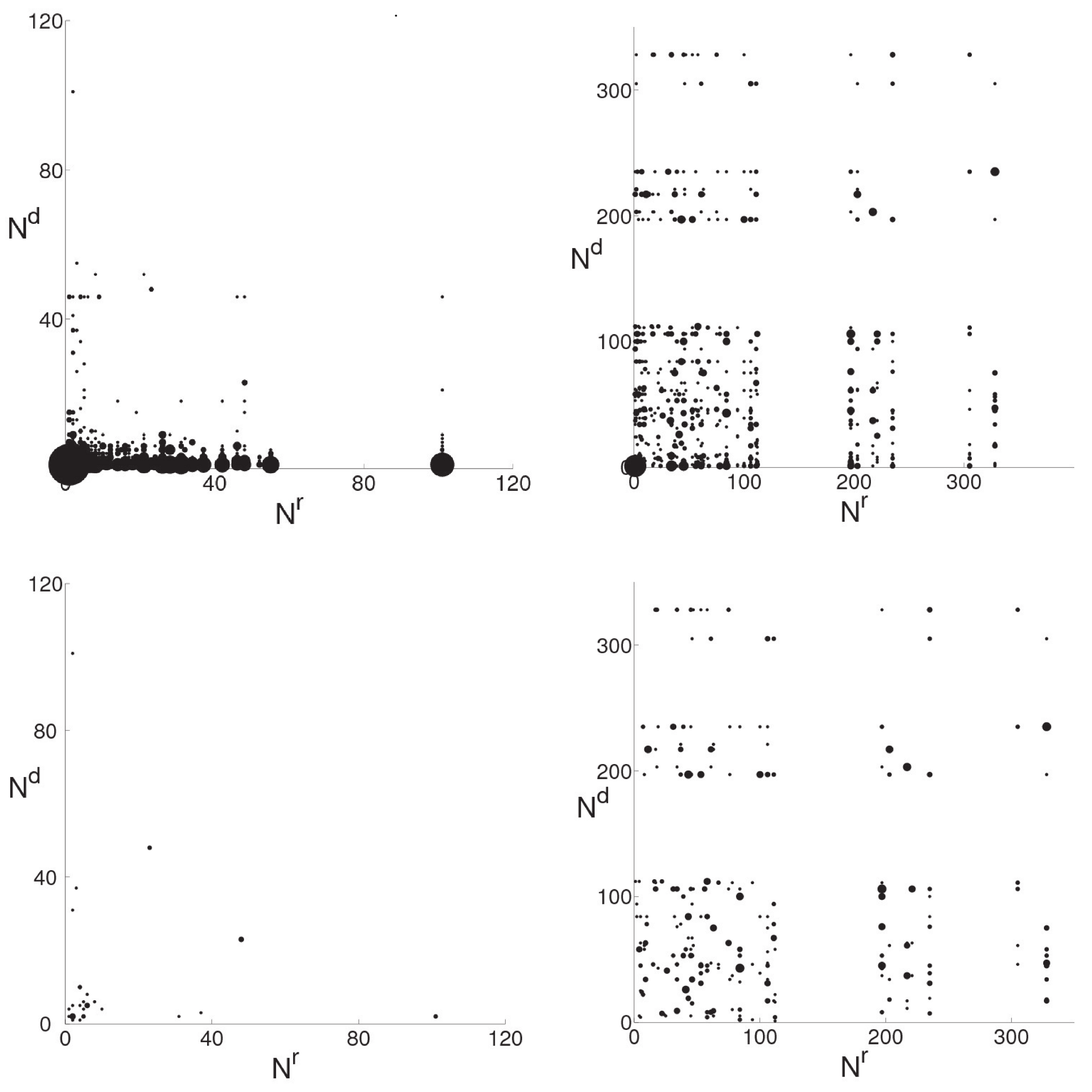}
\caption{{\bf Revert and mutual revert maps of {\tt Benjamin Franklin} (left) and {\tt Israel and the apartheid analogy} (right).}
Diagrams in upper row show the map of all reverts, whereas only mutual reverts are depicted on the diagrams in the lower row.
$N^{\rm r}$ and $N^{\rm d}$ are the number of edits
made by the reverting and reverted editors respectively. Size of the dots is proportional to the number of reverts by the same reverting
and reverted pair of editors.}  \label{maps}
\end{center} \end{figure}

To detect reverts we calculated the MD5 \cite{MD5} hash for each revision, and reverts
were identified by comparing the hash of different revisions.  Let $\dots,
i-1, i, i+1, \dots, j-1, j, j+1, \dots$ be stages in the history of an
article. If the text of revision $j$ coincides with the text of revision
$i-1$, we considered this a revert between the editor of revision $j$ and $i$
respectively.  Let us denote by $N_i$ the total number of edits in the given
article of that user who edited the revision $i$. We characterize reverts by
pairs $(N_i^{\rm d}, N_j^{\rm r})$, where $r$ denotes the editor who makes the
revert, and $d$ refers to the reverted editor (self-reverts are excluded).
Figure~\ref{maps} represents the {\it revert map} of the non-controversial {\tt
  Benjamin Franklin} and the highly controversial {\tt Israel and the
  apartheid analogy} articles. Each mark corresponds to one or more reverts.
The revert maps already distinguish disputed and non-disputed articles, and we
can improve the results by considering only those cases where two editors
revert each other mutually, hereafter called {\it mutual reverts}. This causes
little change in disputed articles (compare the right panels of
Figure~\ref{maps} but has great impact on non-disputed articles (compare left
panels).

\subsubsection*{Controversy measure}

Based on the rank (total edit number within an article) of editors, two main
revert types can be distinguished: when one or both of the editors have few
edits to their credit (these are typically reverts of vandalism since vandals
do not get a chance to achieve a large edit number, as they get banned by
experienced users) and when both editors are experienced (created many edits).  In order to express
this distinction numerically, we use the {\it lesser} of the coordinates $N^{\rm d}$
and $N^{\rm r}$, so that the total count includes vandalism-related reverts as well,
but with a much smaller weight. Thus we define our raw measure of
controversiality as

\begin{equation}
M_{\rm R} = \sum_{(N_i^{\rm d}, N_j^{\rm r})} min(N_i^{\rm d}, N_j^{\rm r})
\end{equation}

Once we developed our first auto-detection algorithm based on $M_{\rm R}$, we
iteratively refined the controversial and the noncontroversial seeds on
multiple languages by manually checking pages scoring very high or very low.
In this process, we improved $M_{\rm R}$ in two ways: first, by multiplying with the
number of editors $E$ who ever reverted mutually  (the larger the armies, the larger the war) and define $M_{\rm I}=E\times M_{\rm R}$
and second, by
censuring the topmost mutually reverting editors (eliminating cases with conflicts between two persons
only). Our final measure of controversiality $M$ is thus defined by

\begin{equation}
M = E \times \sum_{(N_i^{\rm d}, N_j^{\rm r}) < max} min(N_i^{\rm d}, N_j^{\rm r}).
\end{equation}

\subsubsection*{Evaluation and accuracy}

\begin{figure}[!ht]
\begin{center}
 \includegraphics[width=0.5\textwidth]{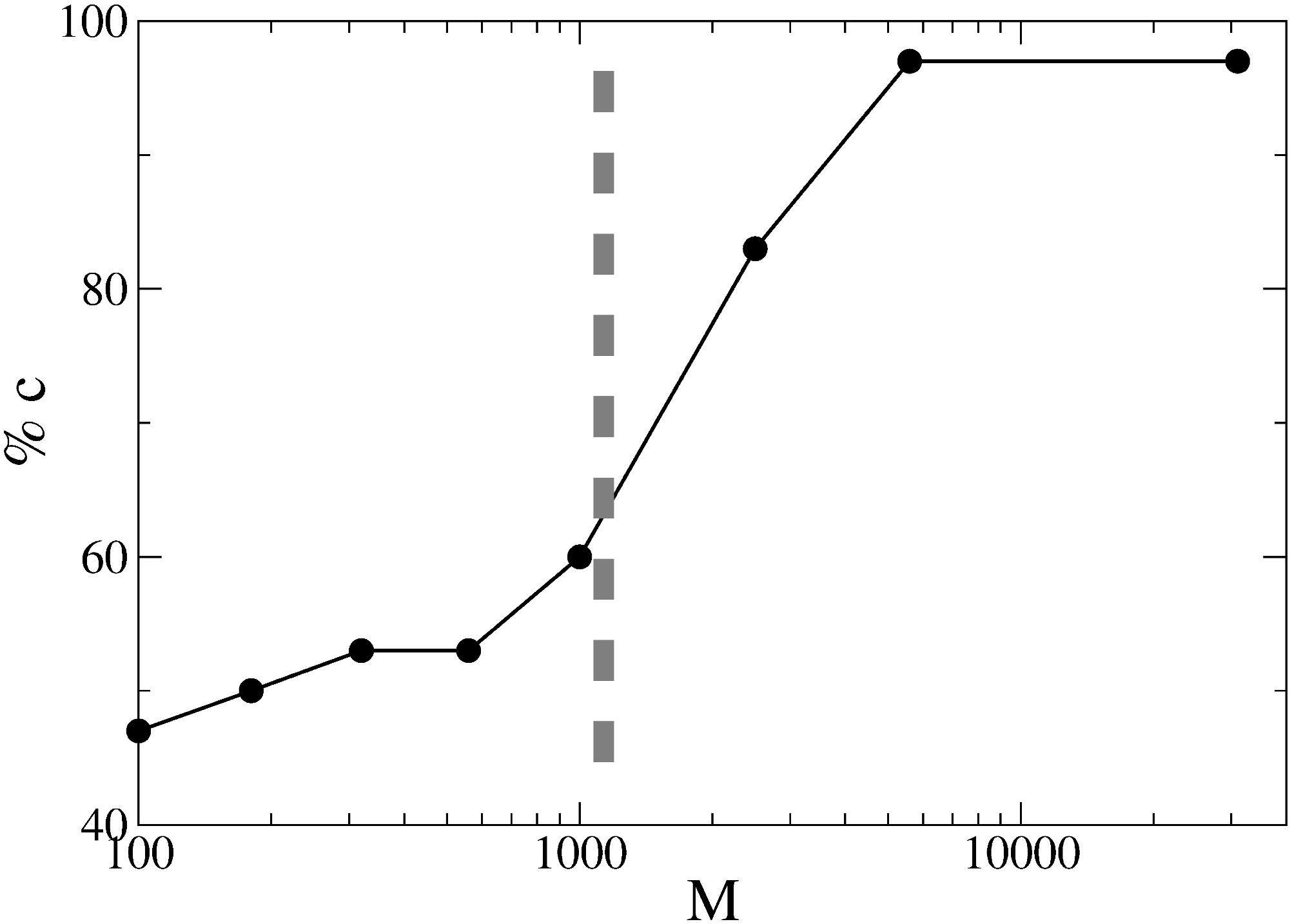}
\caption{{\bf The percentage of true positives in detection of controversial articles compared to human judgment at different values of $M$.}
A threshold of $M\approx 1000$ for controversiality is selected according to this diagram.}\label{c-M}
\end{center}
\end{figure}

One conceptually easy (but in practice very labor-intensive) way to validate
$M$ is by simply taking samples at different $M$ values and counting how many
controversial pages are found (see Figure~\ref{c-M}), considering human judgment as the ``truth''.
We have checked this measure for six different languages and concluded that its overall performance is superior to
other measures \cite{sumi2011b}.

\section*{Results and Discussion}\label{sec:results}
\begin{figure}[!ht]
\begin{center}
 \includegraphics[width=0.5\textwidth]{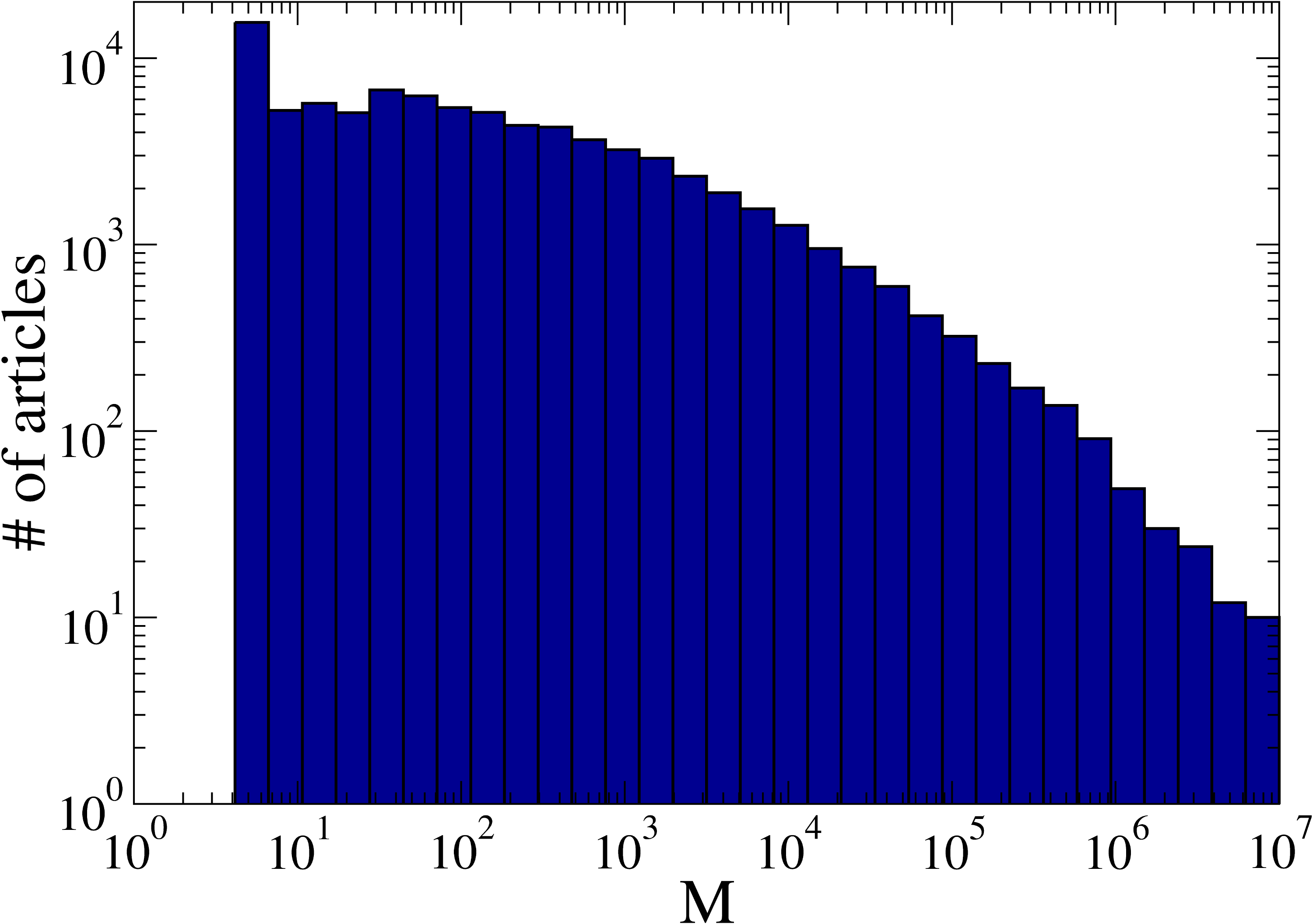}
 \caption{{\bf Histogram of articles according to their controversiality measure $M$.} There are some 84 k articles with $M>10^0$,
12 k controversial articles with $M>10^3$, and less than 100
super-controversial  articles with $M>10^6$}\label{m-dist}
 \end{center} \end{figure}

Having validated the $M$-based selection process, we can start analyzing the
controversial and peaceful articles from a variety of perspectives. We
calculated $M$ for all the articles in the sample -- a histogram is shown in
Figure~\ref{m-dist}.  The primary observation here is that the overall
population of controversial articles is very small compared to the large number
of total articles.  Out of our sample of 233 k articles, there are some 84 k
articles with nonzero $M$, and only about 12 k with $M>10^3$.  The number of
super-controversial articles with $M>10^6$ is less than 100.

We mention in passing that the topical distribution of the controversial
issues differs significantly spatially (across different language editions of
WP): for example, soccer-related issues are massively controversial in the
Spanish WP but not elsewhere. There are flashpoints common to all languages
and cultures, in particular religious and political topics, but we leave the
detailed cross-cultural analysis for another occasion
\cite{yasseri2012b}. Here we focus on the temporal aspects of conflict (based,
unless specifically mentioned otherwise, on the English WP), first at the {\it
  micro-dynamic} level (hours, days, and weeks), and next on a macro timescale
(the lifetime of the article, typically measured in years) to see the {\it
  overall patterns} of conflicts.

\subsection*{Micro-dynamics of conflicts}

\begin{figure}[!ht]
\begin{center}
 \includegraphics[width=0.5\textwidth]{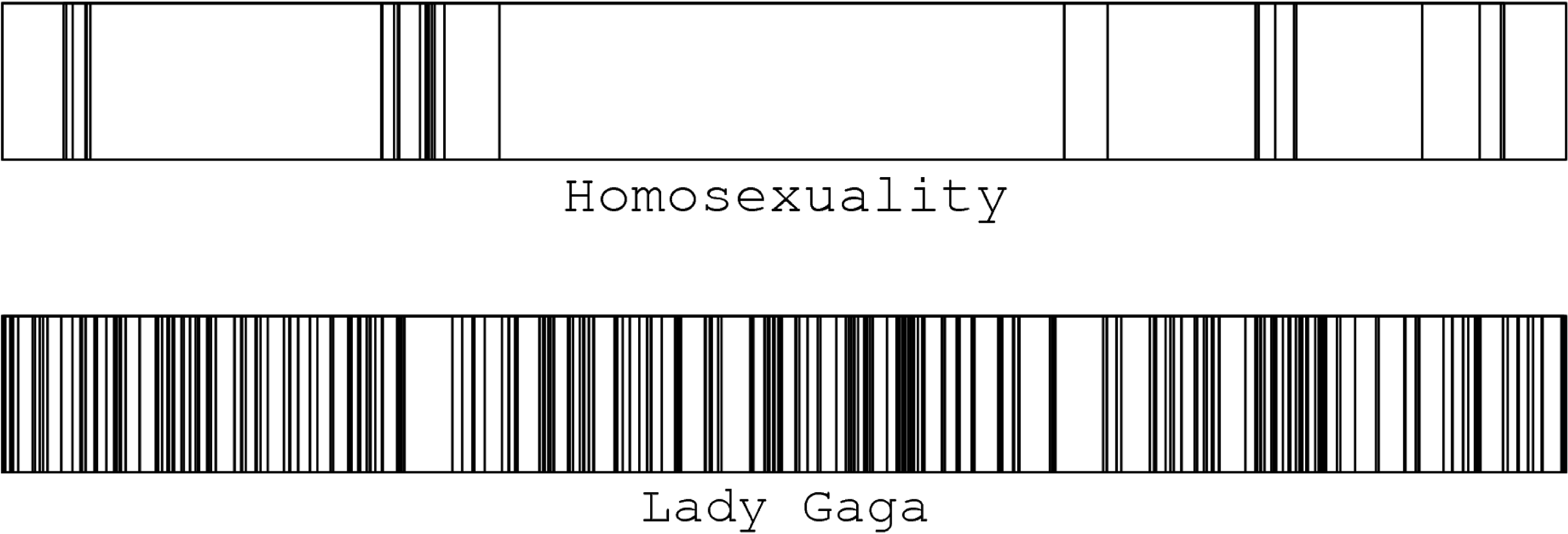}
 \caption{{\bf Temporal edit patterns of {\tt Lady Gaga} and {\tt Homosexuality}
     during a one month period (12/2009).} The horizontal axis is time, each
   vertical line represents a single edit. Despite the large differences in
   average time intervals between successive edits, the bursty editing
   pattern is common to both cases.}\label{gaga}
 \end{center} \end{figure}

Once we have a reliable measure of controversiality, not only can we find and
rank controversial issues in WPs, but we actually begin to see important
phenomena and common characteristics of wars and disputes. Here we report our
findings on the temporal characteristics of edits on high and low
controversiality pages.  We make use of the fact that in the WP dump a
timestamp with one second precision is assigned to each edit. One month of
activity (the time-line of all edits irrespective of who performed them) on
two sample articles are depicted in Figure~\ref{gaga}.

West et. al.\cite{west2010} and Adler et. al. \cite{adler2011} have developed
vandalism detection methods based on temporal patterns of edits. In both
studies the main assumption is that offensive edits are reverted much faster
than normal edits, and therefore, by considering the time interval between an
arbitrary edit and its subsequent reverts, one can classify vandalized versions
with high precision.

\subsubsection*{Edit frequency}
\begin{figure}[!ht]
\begin{center}
 \includegraphics[width=0.5\textwidth]{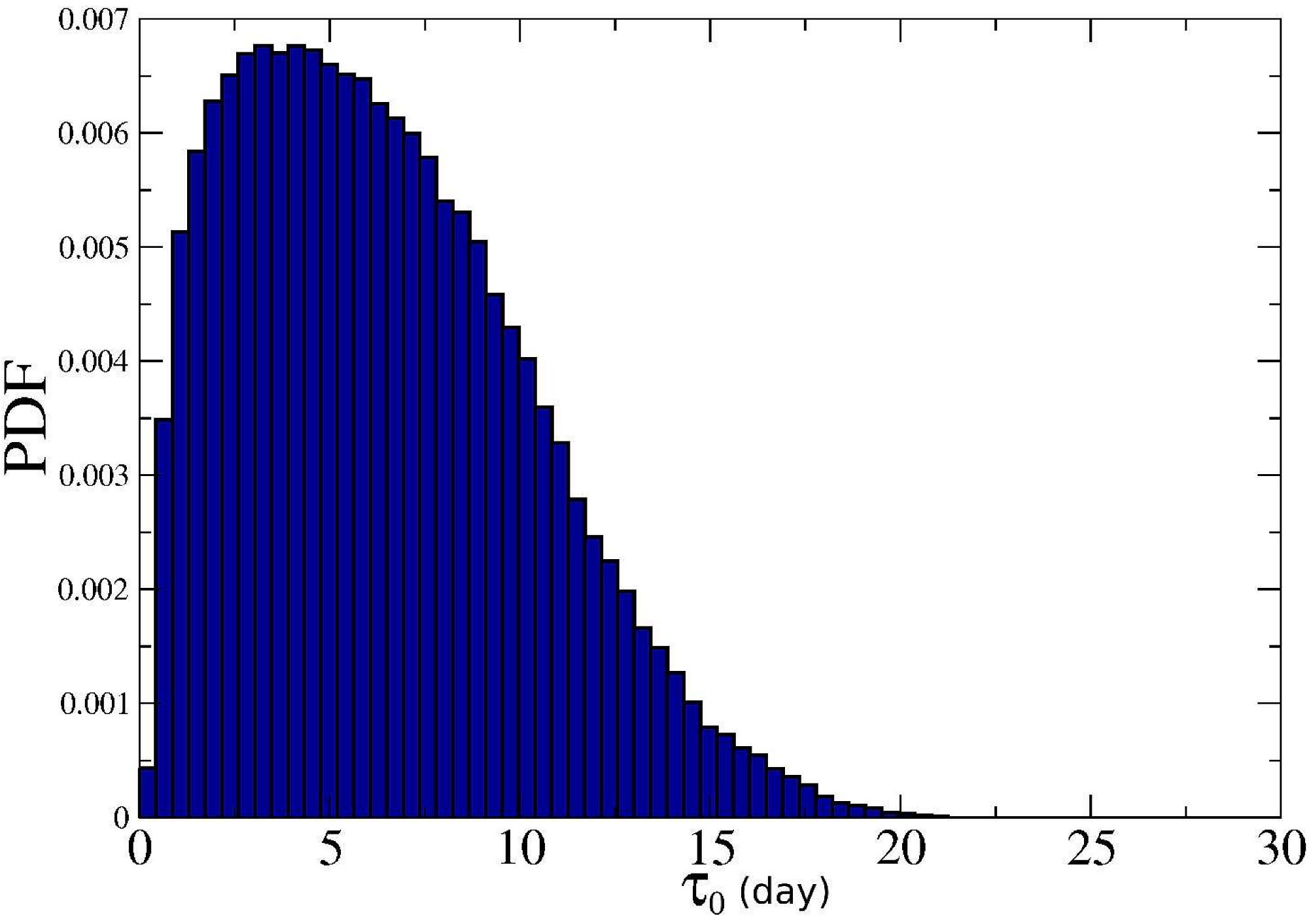}
 \caption{{\bf PDF of the average time $\tau_0$ between two successive edits
     of articles measured in days.} In any two week period most of the
   articles are edited twice or more.}\label{p-tau}
 \end{center} \end{figure}

\begin{figure}[!ht]
\begin{center}
 \includegraphics[width=0.5\textwidth]{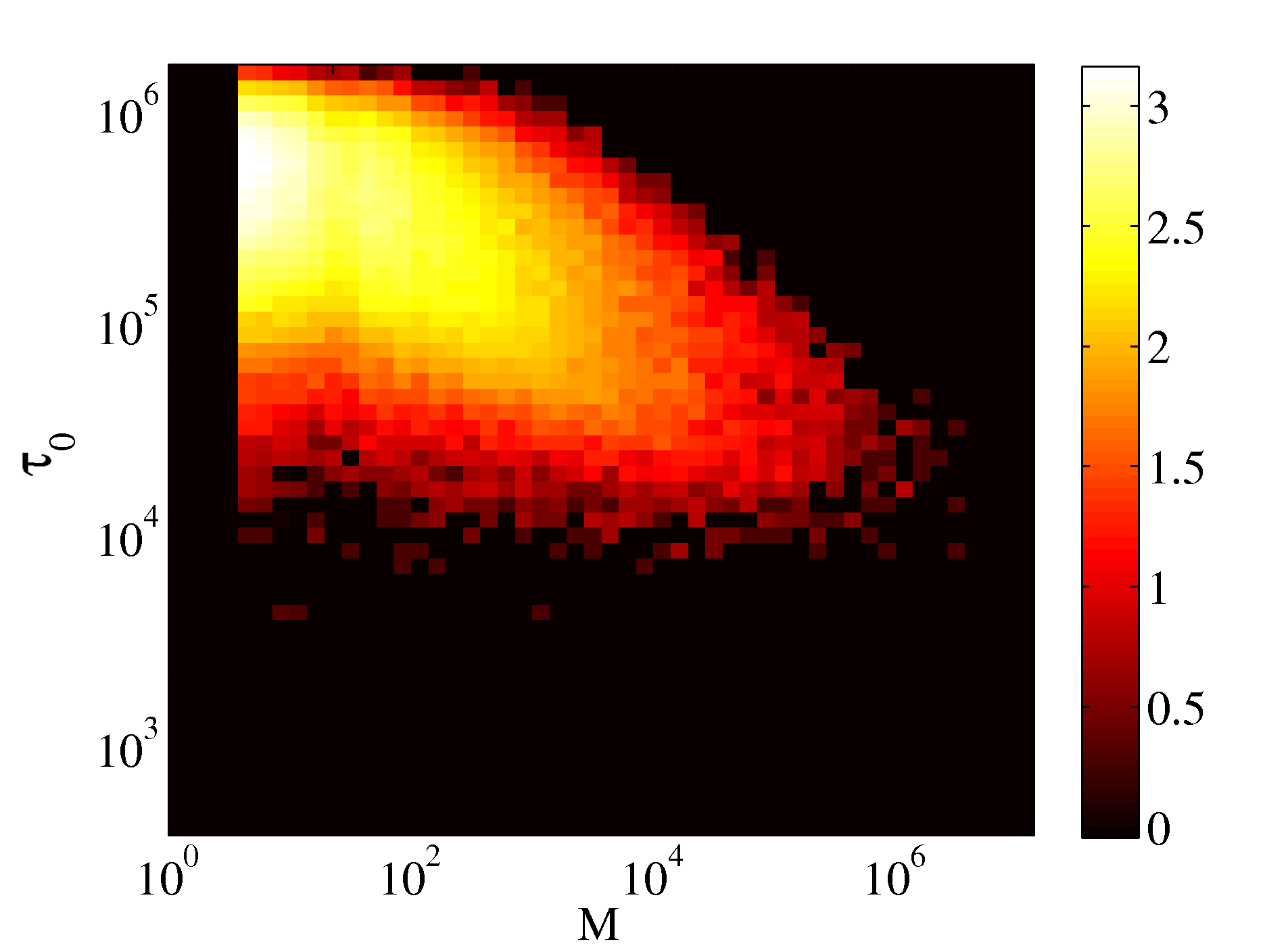}
\caption{{\bf Scatter plot of the average time interval between successive
edits and the controversy measure.} Color coding is according to logarithm of the density of points. The correlation coefficient $C=-0.03$.}\label{M-tt}
\end{center}
\end{figure}

Most of the articles are frequently edited. Figure~\ref{p-tau} shows the
empirical probability density function of the average time $\tau$ between
two successive edits. As already noted in \cite{ratkiewicz2010} edit frequency also depends on
the controversiality of a page, and one expects higher edit frequency for more
controversial pages. However, as Figure~\ref{M-tt} makes clear, the correlation
is quite weak (correlation coefficient $C=-0.03$).

\subsubsection*{Burstiness}

\begin{figure}[!ht]
\begin{center}
 \includegraphics[width=0.5\textwidth]{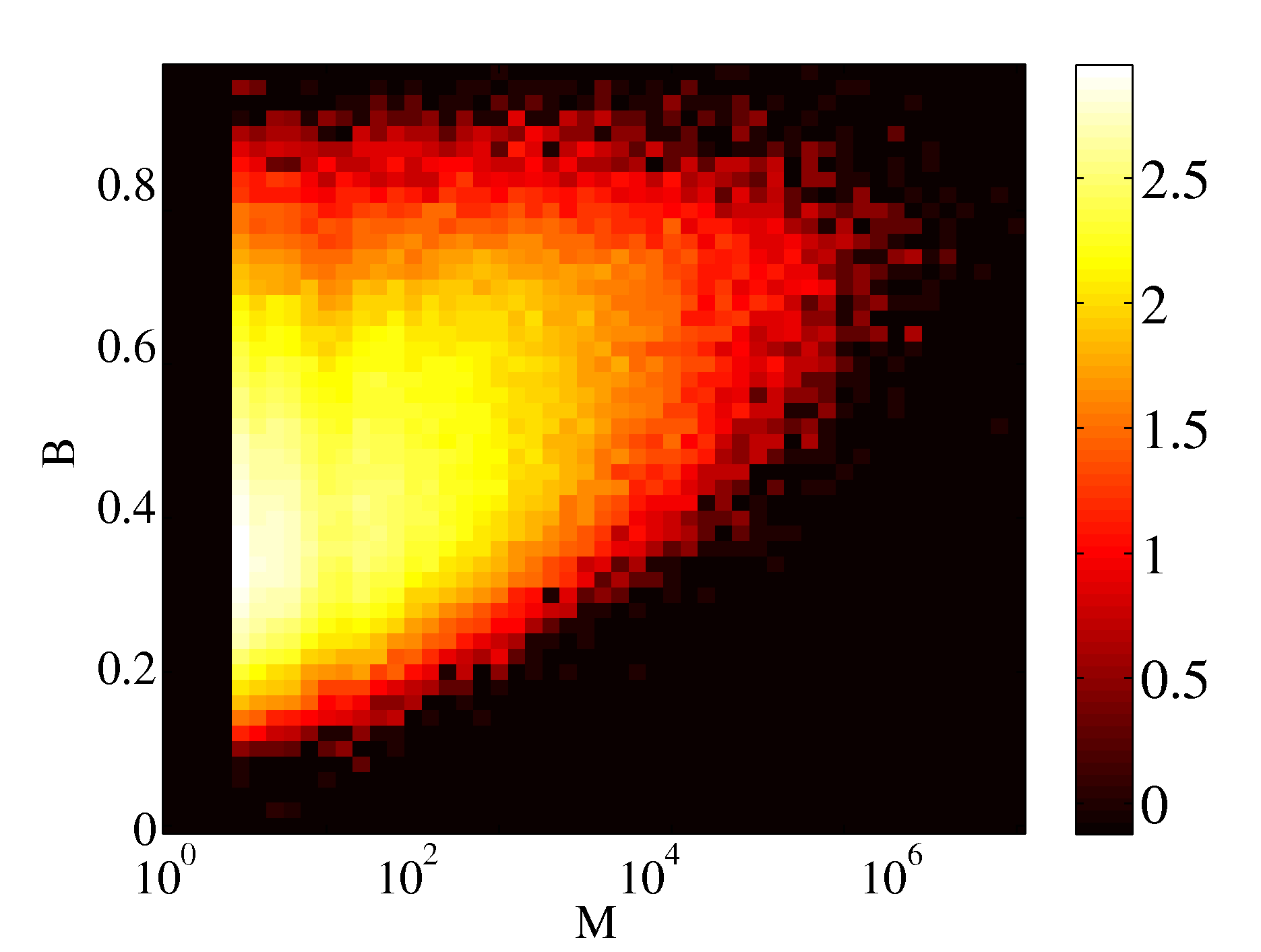}
\caption{{\bf Scatter plot of burstiness and the controversy measure.} Color coding according to logarithm of the density of points.
The correlation coefficient $C=0.05$.}\label{M-B}
\end{center}
\end{figure}

It is clear that edits are clustered in a way that there are many edits done
in a rather short period, followed by a rather long period of silence.  This
feature is known in the literature as {\it burstiness}
\cite{goh2008,barabasi2005}, and is quantified based on the coefficient of
variation by a simple formula as

\begin{equation}
 B \equiv \frac{\sigma_{\tau} - m_{\tau}}{\sigma_{\tau} + m_{\tau}}
\end{equation}
where $m_{\tau}$ and $\sigma_{\tau}$
denote respectively the mean and standard deviation of the interval $\tau$
between successive edits. We have calculated $B$ for all the articles in the sample, considering
all the edits made on them by any user. As it can be seen from Figure \ref{M-B}, overall
burstiness of edits correlates rather weakly with controversiality ($C=0.05$).
   
\begin{figure}[!ht]
 \begin{center}
 \includegraphics[width=0.45\textwidth]{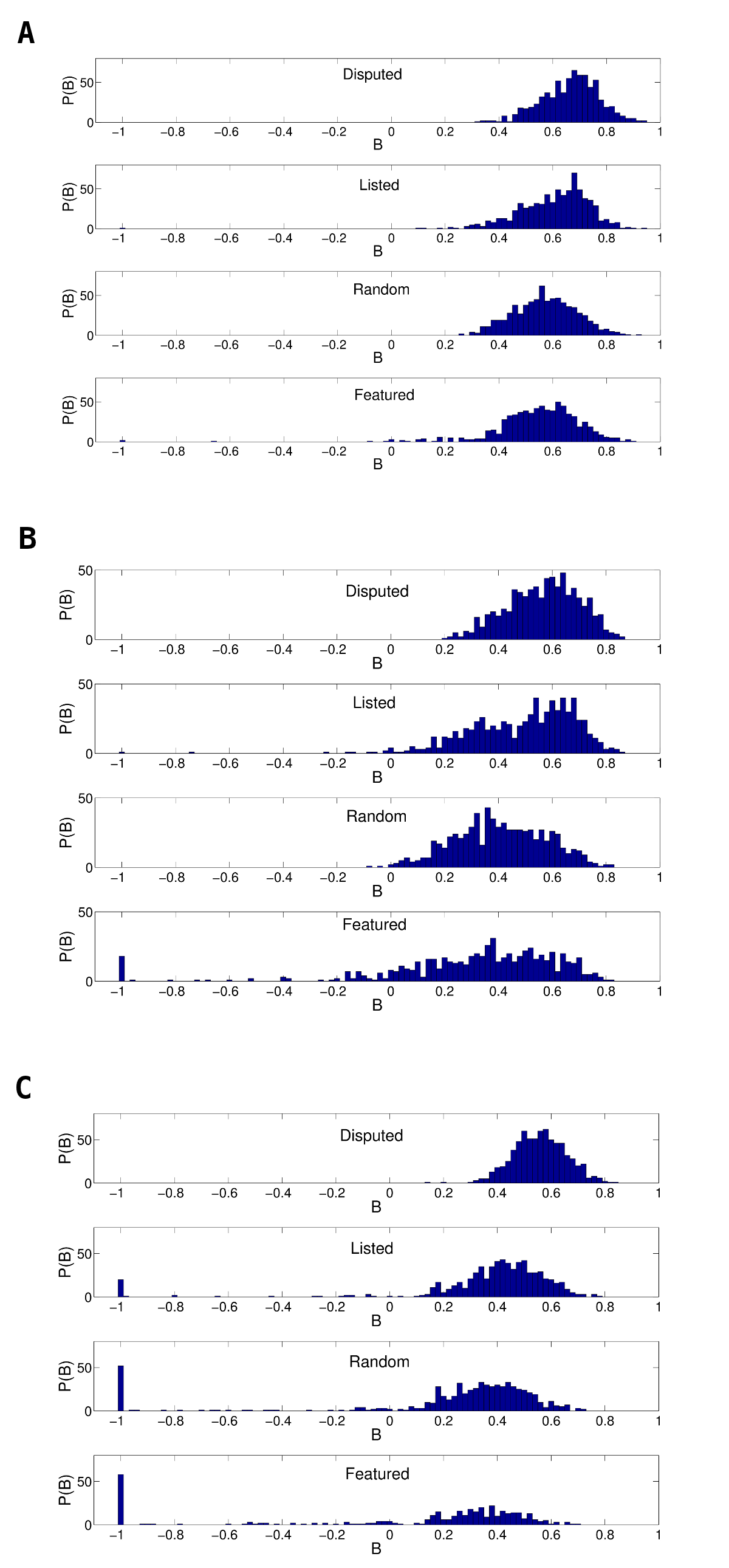}
 \caption{{\bf Histogram of burstiness of A) all edits, B) reverts, and C) mutual reverts for four classes of articles.}
High controversy ($M >1000$, topmost panels),listed as controversial (2nd panels), randomly
selected (3rd panels), and featured articles (bottom panels).}\label{bhist}
 \end{center}
\end{figure}

To see the impact of controversiality on burstiness we calculated $B$ for
different groups of articles separately: {\it Disputed} articles $(M > 1000)$,
{\it Listed} articles coming from the {\tt List of controversial articles} in WP \cite{wiki-list},
{\it Random}ly selected articles, and {\it Featured} articles (assumed to be least
controversial given WPs stringent selection criteria for featuring an
article).  The histograms in Figure~\ref{bhist}(A) show the PDF of $B$ in these
four classes.
As can be seen, the peaks are shifted to the right (higher $B$) for more
controversial articles, but not strongly enough to base the detection
of controversy on burstiness of editorial activity alone. Reverting is
a useful tool to restore vandalized articles, but it is also a popular weapon
in heated debates. Figure~\ref{bhist}(B) shows the distribution of $B$ calculated not
for all edits, but for reverts alone: the shift is now more marked.
Finally, we considered an even stronger form of
warfare: {\it mutual reverts}. It is evident that the temporal pattern of
mutual reverts provides a better characterization of controversiality than
that of all edits or all reverts, and the very visible shift observed in
Figure~\ref{bhist}(C) constitutes another, albeit less direct, justification of
our decision to make mutual reverts the central element in our measure of
controversiality.

\begin{figure}[!ht]
\begin{center}
 \includegraphics[width=0.75\textwidth]{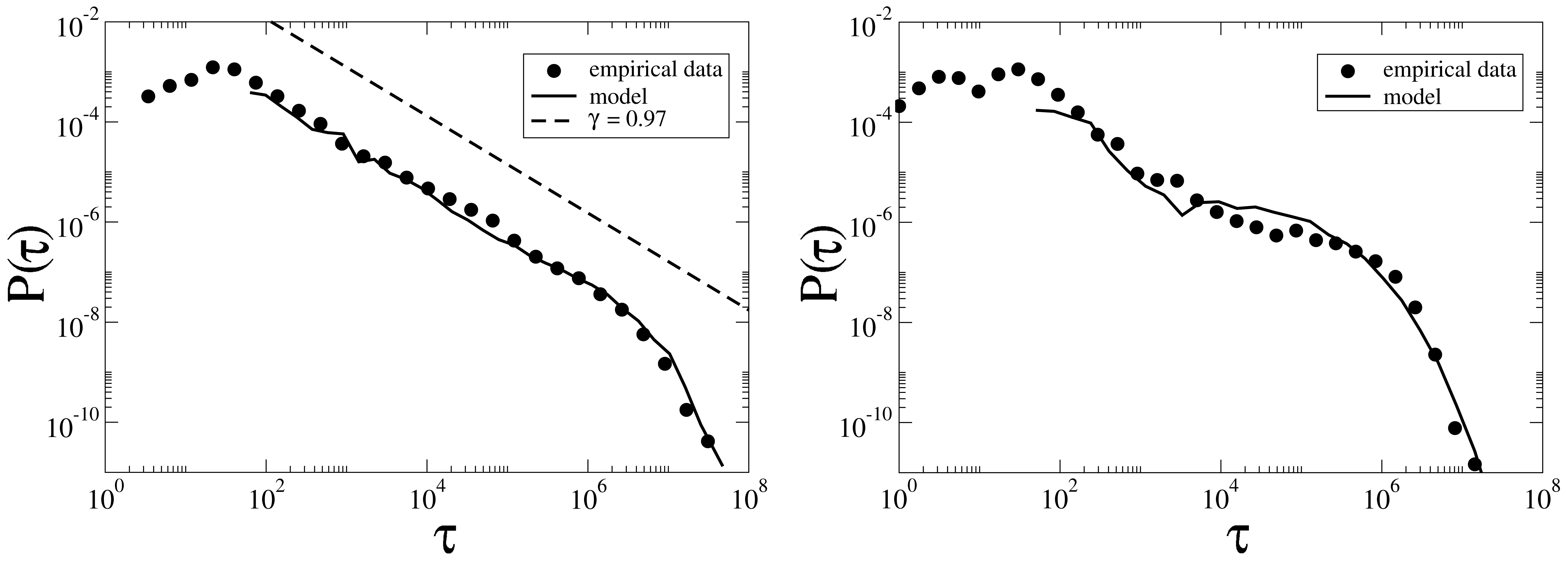}
\caption{{\bf PDF of intervals between two successive edits on an article (in
    seconds) for two samples of highly/weakly disputed articles (left/right
    panel).}  Each sample contains 20 articles and the average $\tau$ for all
  articles is about 10 hours.  Circles are empirical data and solid lines are
  model fit, with values $L=20$, $P=0.9$ and $L=500$, $P=0.5$ respectively for
  disputed and non-disputed samples. The dashed line in the left panel is the
  power law with exponent  $\gamma=0.97$.}\label{tau} \end{center}
 \end{figure}

To gain a better understanding of the microdynamics of edit wars, we
  selected two samples of 20 articles each, extracted from a pool of articles
  with average successive edit time intervals of 10 hours~$\pm5\%$. One sample
  contains the most controversial articles in the pool with
  $10^4<M<7\times10^4$, whereas the other one contains the most peaceful
  articles with $100<M<150$.  The probability distribution of time $\tau$
between edits for these samples is shown in Figure~\ref{tau}. Both samples have
a rather fat-tailed distribution with a shoulder in the distribution (as
observed both in the empirical data and the model calculation), indicating
that a characteristic time, $\tau \approx 10^5$ seconds (one day), is present
in the system. However, only the sample consisting of controversial articles
displays a clear power-law distribution, $P(\tau)\sim \tau^{-\gamma}$,with
$\gamma=0.97$.  All exponents were calculated by applying the
  Gnuplot implementation \cite{gnuplot} of the
  nonlinear least-squares Marquardt-Levenberg algorithm \cite{wiki:levenberg-marquardt} on the log-binned data
  with an upper cut-off to avoid system size effects.

To fit the data depicted in Figure~\ref{tau}, we
used a model based on a queuing mechanism introduced in \cite{barabasi2005}
and further developed in \cite{vazquez2006}. Here we briefly explain its basis
and how we use it to model our empirical findings.  Let us assume that there
is a list of $L$ articles and there is only one editor (mean-field
approximation) who edits at each step once.  With probability $1-P$, the
editor selects the article to edit from the list randomly and with no
preference among $L$ choices. With probability $P$ the articles will be
selected according to a priority $x_i\in (0,1) $ which is assigned randomly to
the $ith$ article after each edit on it. The key parameters are $L,P$ and the
real time $t_p$ associated to the model time step.  Controversial articles are
fitted well by $P$ close to 1 and small $L$.  Uncontroversial articles fit
with large $L$ and smaller $P$, in nice agreement with the real situation,
where editors tend to edit a few controversial articles more intentionally and
many peaceful articles in a more or less uncorrelated manner with no bias and
memory.
To check the validity of the model, we calculated the ratio of the number of
controversial articles (with $M > 1000$) to the rest of the articles ($M <
1000$) to be $\sim 0.052$, which is in nice agreement with the fitting model
parameters, 20/500=0.04.

\begin{figure}[!ht]
\begin{center}
 \includegraphics[width=0.75\textwidth]{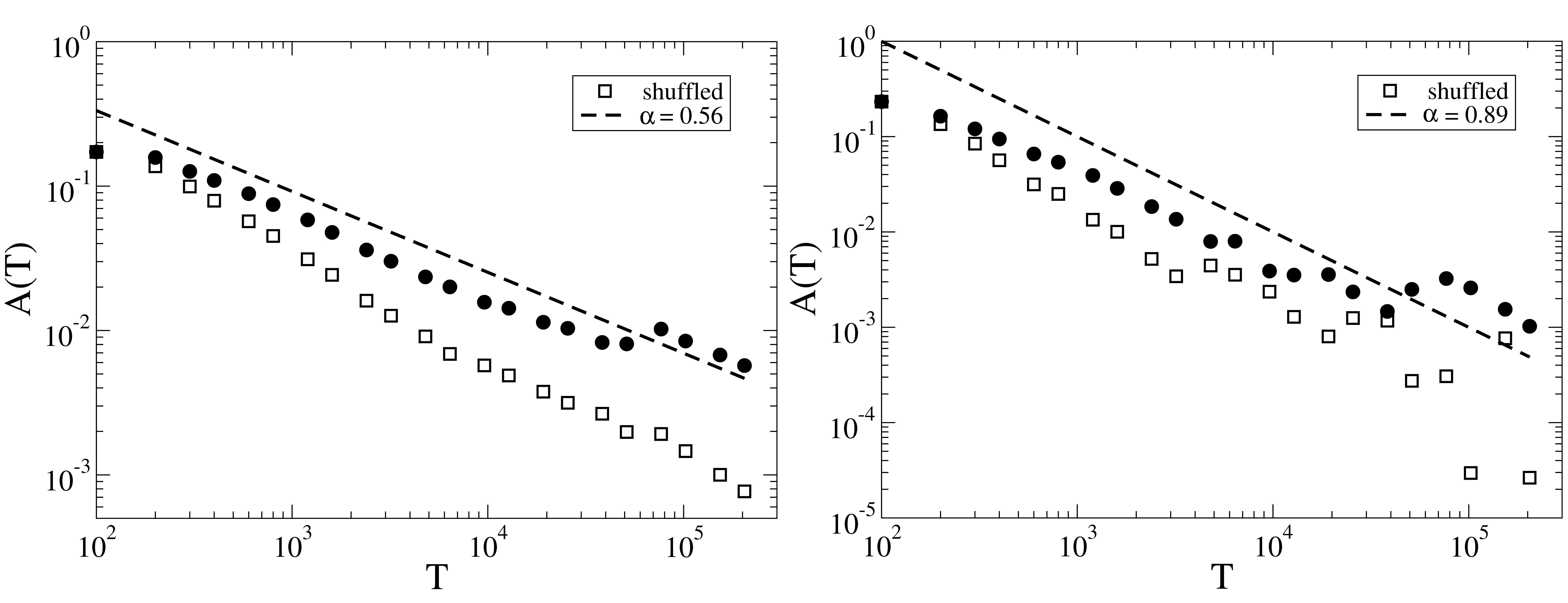}
\caption{{\bf Autocorrelation function of edits sequences for two samples of highly/weakly
disputed articles (left/right panel).} Circles are for the original sequences,
  empty squares correspond to the shuffled sequences. Dashed lines are power-law fits.}\label{A}
\end{center}
\end{figure}

Another important characteristic quantity is the autocorrelation function
$A(T)$. To calculate it, first we produce a binary series of 0/1 $X(t)$
similar to the one in Figure~\ref{gaga}. Then $A(T)$ is computed simply as
\begin{equation}
 A(T)= \frac{\langle X(t)X(t+T)\rangle_t - \langle X(t) \rangle^2_t}{\langle X(t)\rangle_t-\langle X(t)\rangle^2_t}
\end{equation}
where $\langle . \rangle_t$ stands for the time average over the whole
series. $A(T)$ for the same samples of controversial and peaceful articles
are shown in Figure~\ref{A}. We calculate the same quantity for a shuffled sequence of events as a reference. The shuffled sequence has
the same time interval distribution as the original sequence, but with a randomized order in the occurrence of events. In both cases, a power-law of $A(T)\sim T^{-\alpha}$
describes $A(T)$ very well.  Usually it is assumed that slow (power law) decay
of the autocorrelation function is an indicator for long time memory
processes.  However, if {\it independent} random intervals taken from a power
law distribution separate the events, the resulting autocorrelation will also
show power law time dependence \cite{vajna2012,karsai-up}.  Assuming that the
exponent of the independent inter-event time distribution is $\alpha$ and the
exponent of the decay of the correlation function is $\gamma$, we have the
relationship $\alpha + \gamma = 2 $. Deviations from this scaling law reflect
intrinsic correlations in the events.

\begin{figure}[!ht]
\begin{center}
 \includegraphics[width=0.75\textwidth]{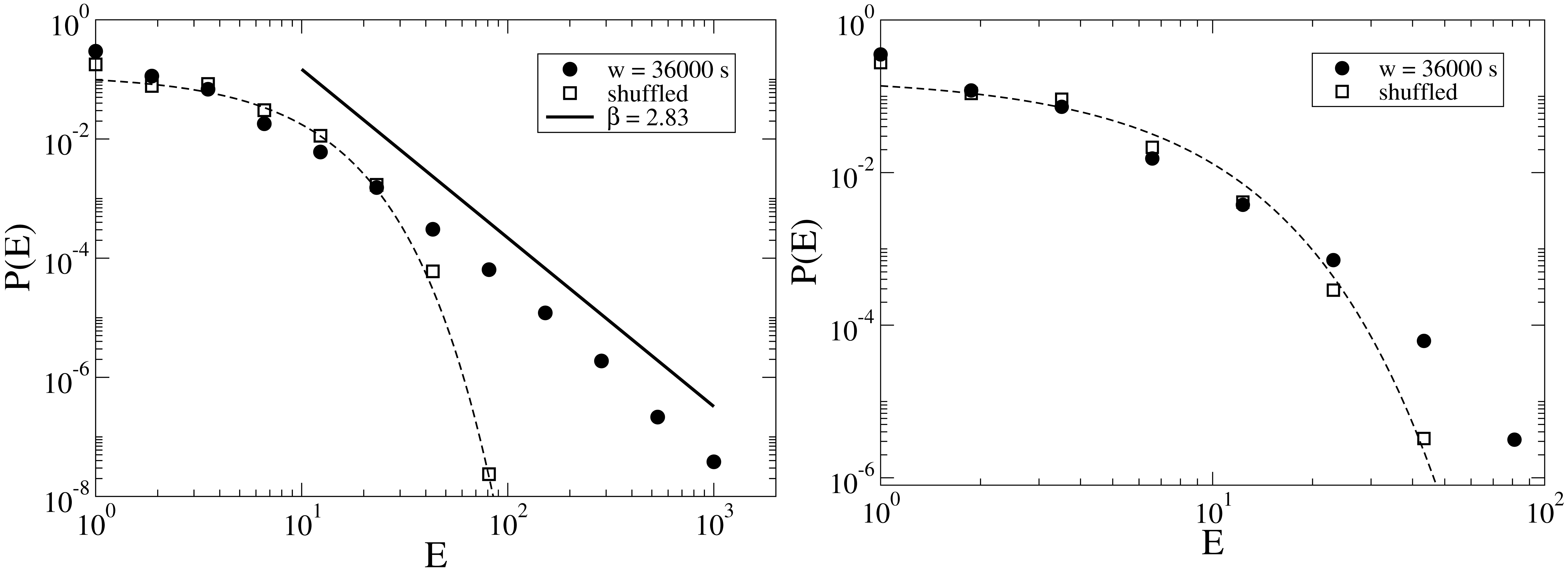}
\caption{{\bf Distribution of $E$ for two samples of highly/weakly
disputed articles (left/right panel).} Circles are for the original sequences, whereas empty squares correspond to the shuffled sequences. Dashed lines
are exponential fits to the $P(E)$ for shuffled data and solid line
in the left panel is a power-law with $\beta=2.83$.}\label{E}
\end{center}
\end{figure}

There is another measure which indicates long time correlations
between the events even more sensitively. Take a period to be bursty if the
time interval between each pair of successive edits is not larger than $w$,
and define $E$ as the number of events in the bursty periods.  If events in
the time series are independent and there is no memory in the system (i.e. in
a Poisson process), one can easily show that $P(E)$ should have an exponential
decay, whereas in the presence of long range memory, the decay is in the form
of $P(E)\sim E^{-\beta}$ \cite{karsai-up}.  In Figure~\ref{E}, $P(E)$ is shown
for samples of highly controversial and peaceful articles. In the high
controversy sample a well defined slope of -2.83 is observed, while in the low
controversy sample edits are more independent and $P(E)$ is very close to the
one obtained for the shuffled sequence. Note that by shuffling the
sequence of time intervals, all the correlations are eliminated and the
resulting sequence should mimic the features of an uncorrelated occurrence
of the intervals.

The same measurements are performed for a sample of users, see Figure~\ref{user-burst} in Supporting
Information. In Table.\ref{tab:exponents}, a summary of the scaling exponents
for the both article samples and users is reported.

The simplest explanation of these results is to say that conflicts induce
correlations in the editing history of articles. This can already be seen in
Figure~\ref{A}, where shuffling influences the decay of the autocorrelation
functions much more for high-$M$ articles than for low-$M$ ones. For the more
sensitive measure $P(E)$ the original and the shuffled data are again quite
close to each other for the low-$M$ case, while a power-law type decay can be
observed in the empirical data for high-$M$ articles.

\subsection*{Overall patterns of conflicts}

\begin{figure}[!ht]
\begin{center}
\includegraphics[width=0.5\textwidth]{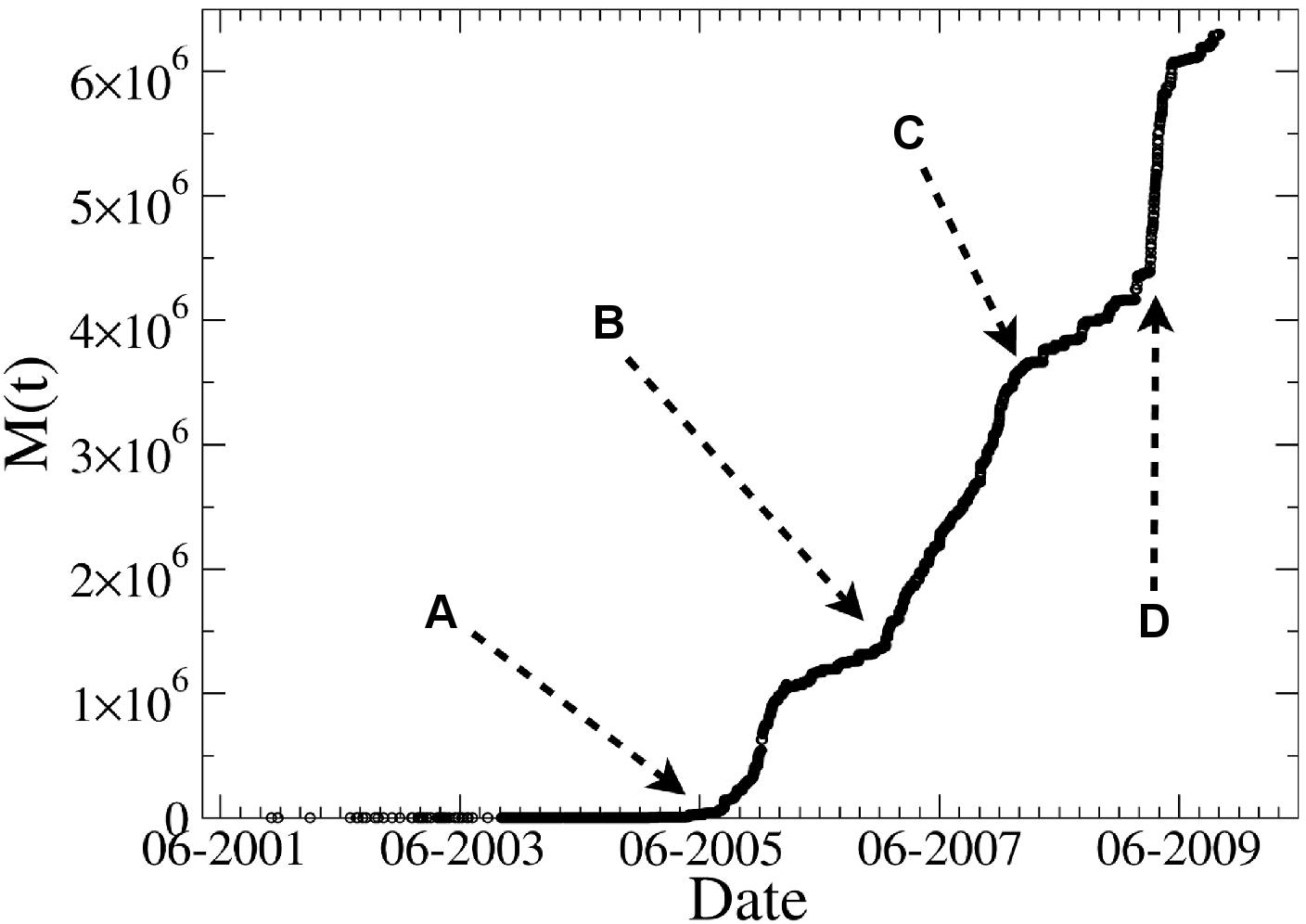}
\caption{{\bf Time evolution of the controversy measure of {\tt Michael
    Jackson}.} A: Jackson is acquitted on all counts after five month
  trial. B: Jackson makes his first public appearance since the trial to
  accept eight records from the Guinness World Records in London, including
  {\it Most Successful Entertainer of All Time.} C: Jackson issues {\it Thriller
    25}. D: Jackson dies in Los Angeles.}\label{mjackson}
\end{center}
\end{figure}

Before we can consider the macro-scale evolution of $M$ (during the entire
life of the article), we need to make an important distinction between endo-
and exogenous causes of conflict. Our principal interest is with endogenous
forces, which originate in internal sources of conflict and disagreement, but it cannot be denied that in a
significant number of cases conflicts are occasioned by some exogenous event,
typically some recent development related to the real-world subject of the
article rather than to its text (see Figure\ref{mjackson} for some examples).

\subsubsection*{Categorization}

In the presence of significant exogenous events one can best follow the
increase of $M$ as a function of time $M(t)$, but if endogenous edits dominate
(as is the case with most science articles and bibliographies of persons long
dead) it is more natural to trace $M$ as a function of the number of edits on
the article $M(n)$ because temporal frequency of edits changes from time to
time and from article to article, due to many different known and unknown
causes \cite{ratkiewicz2010b,yasseri2012a}. Since exogenous factors are completely
unpredictable, in the following section we try to categorize articles
according to $M(n)$.

Even if we restrict attention to endogenous growth, very different patterns
can be observed in the evolution of $M$, depending not just on the current
controversiality of a subject (by definition, $M$ never decreases except for
small truncation effects due to changes in who are the most engaged pair of
reverters), but also on the micro-dynamics of edit wars. Here we try to
recognize some general features based on numerical properties of $M(n)$ and
its derivative, and categorize the articles accordingly.  We applied a maximum
detection tool to the smoothed derivative curve to locate both the hot periods
of wars and the `consensus reached' situations where the derivative of $M(n)$
is very small or zero. Based on the statistics of the war and consensus
periods, we categorize articles into three main categories.

\begin{figure}[!ht]
\begin{center}
\includegraphics[width=0.5\textwidth]{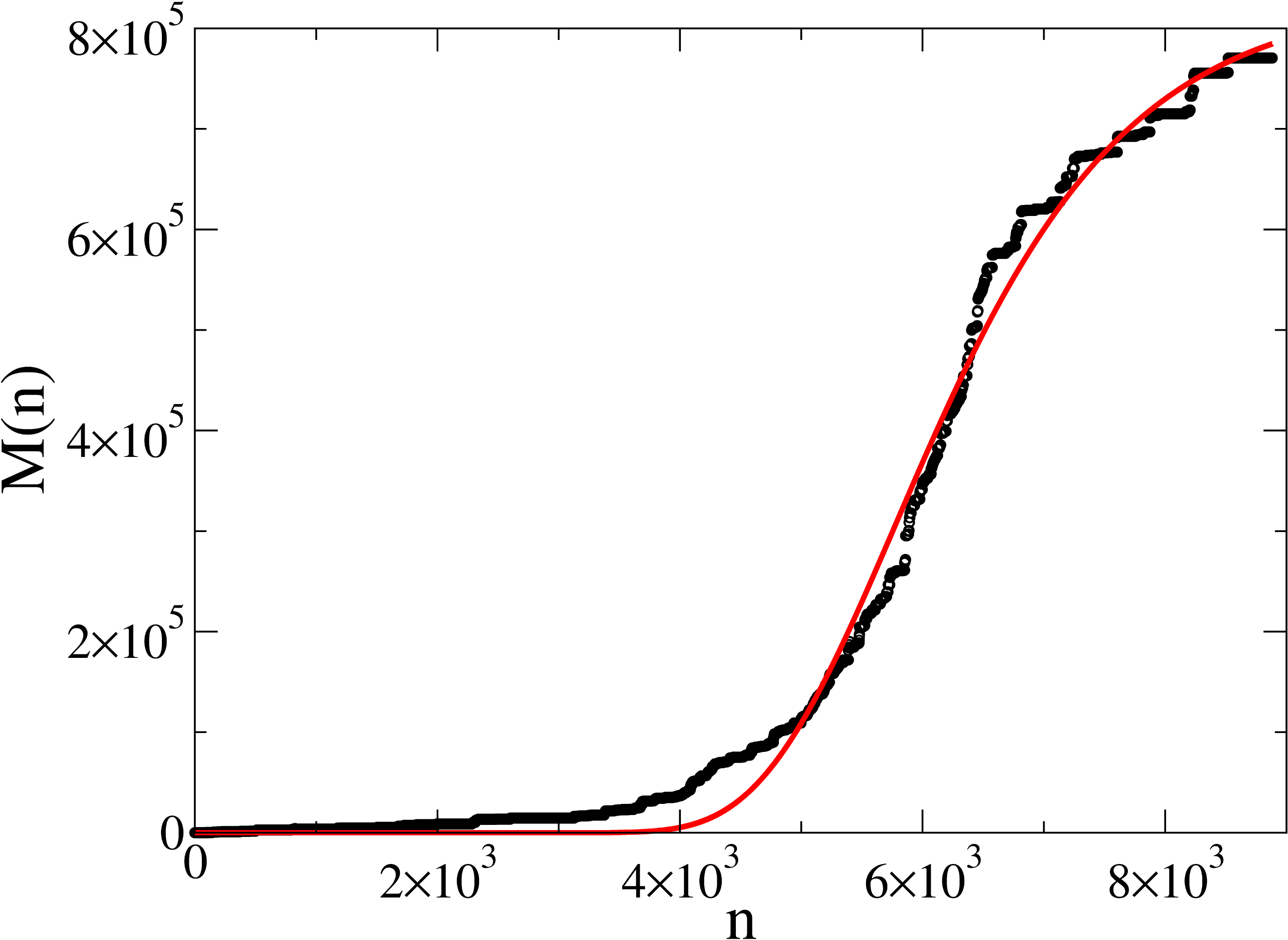}
\caption{{\bf Evolution of controversy measure with number of edits of
    {\tt Jyllands-Posten Muhammad cartoons controversy}, with Gompertz fit shown
    in red.} The initial rapid growth in $M$ tends to saturate, corresponding
  to the reaching to consensus.}\label{catA}
\end{center}
\end{figure}

 {\bf a) Consensus.} The common scenario for the cases where at
the end consensus is reached is the following.  Usually growth starts slowly
and with an increasing acceleration until it reaches a maximum speed of
growth. Afterwards, when the hot period of war is passed, the growth rate
decreases and consensus is reached, where $M$ does not, or only very slightly,
increases upon the next edits. We do not offer a mathematical model for such
growth here, but we note that a Gompertz function
$M(n)=M_{\infty}e^{b{e^{cn}}}$ (with $M_{\infty}$ being the final value of
$M$, and $(b,c<0)$ are the displacement parameter and growth rate
respectively) offers a reasonable fit ($R^2>0.95$) for almost all $M(n)$ in
this category (see Figure~\ref{catA} for an example). In general, the Gompertz
function fares better than sigmoid because it does not force symmetry around
the initial and the final asymptote, and the literature such as
\cite{laird1964} suggests it is a more appropriate model for growth in a
confined space. We leave the matter of how controversiality becomes a
consumable resource for future research, but we find it quite plausible that
certain articles can become so well polished that it becomes extremely hard to
pick a fight about them.

\begin{figure}[!ht]
\begin{center}
\includegraphics[width=0.5\textwidth]{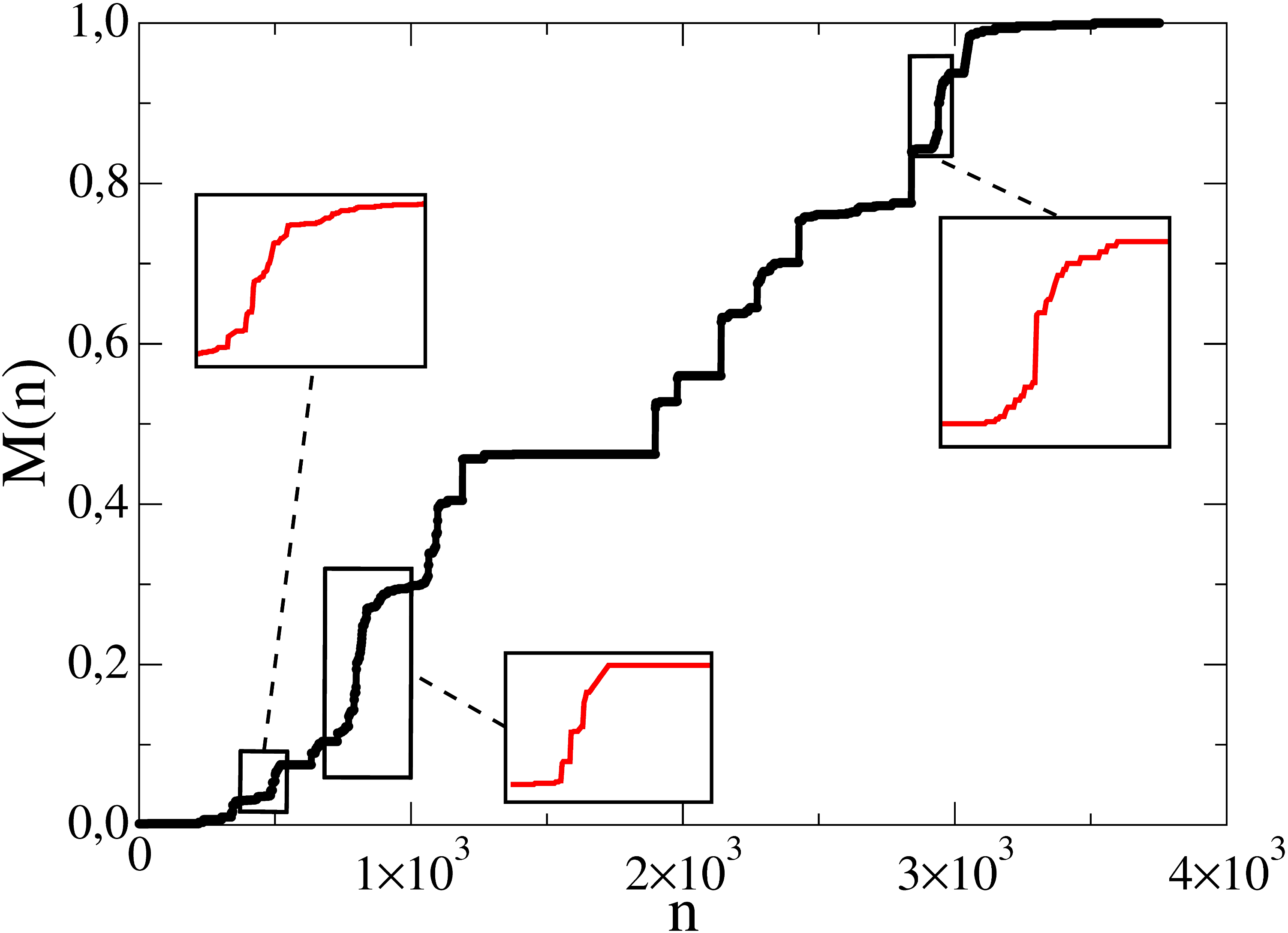}
\caption{{\bf Evolution of controversy measure with number of edits of of
    {\tt Iran} -- the insets depict focuses of some of the local war periods.}
$M(n)$ is normalized to the final value $M_{\infty}$. Cycles of peace and war appear consequently, activated by internal and external causes.}\label{catB}
\end{center}
\end{figure}

{\bf b) Sequence of temporary consensuses}. The common feature
of the articles in this category is sequential appearance of war and peace
periods in a quasi-periodic manner. After the first cycle of war and consensus
as described in (a), internal or external causes initiate another
cycle. Exogenous changes happen completely randomly, but the endogenous
causes may be contributed by a simple mechanism such as a constant influx of
new editors, who are not satisfied with the previously settled state of the
article (see Figure~\ref{catB} for an example).

\begin{figure}[!ht]
\begin{center}
 \includegraphics[width=0.30\textwidth]{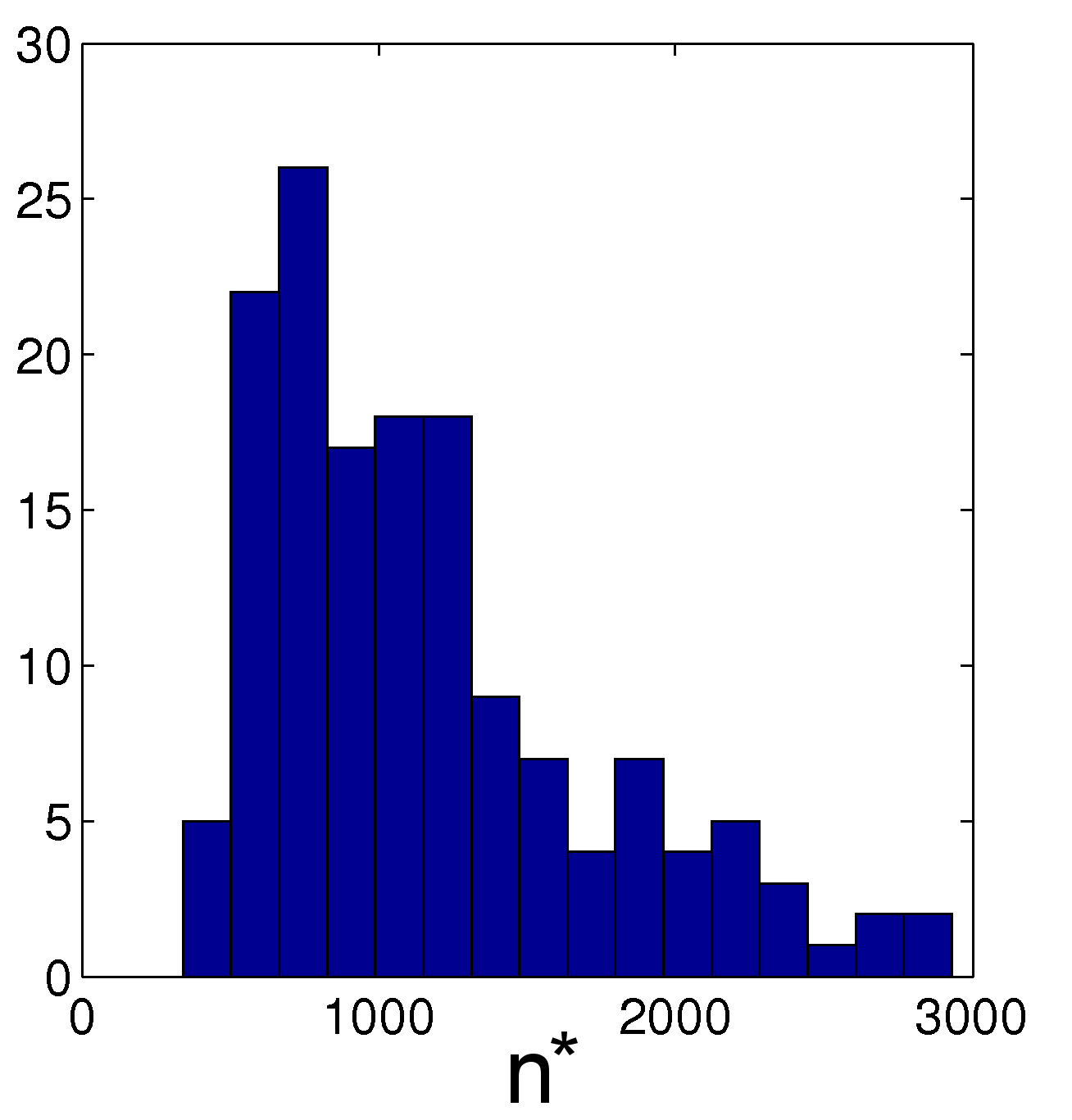}
\caption{{\bf Length of peacful periods.} Histogram of number of edits between two successive war periods for a selected sample of 44 articles which are not driven by
 external events. The average value of $n^*$ is 1300 edits. }
\label{nstar} \end{center} \end{figure}

\noindent
We do not have the means to make the required systematic distinction between
internal end external causes (manual evaluation is too expensive,
auto-detection would require too much world knowledge). Therefore, we created a
limited sample of 44 articles, which are entirely about solid concepts and
facts, in order to measure the periodicity of endogenous controversies.
Figure~\ref{nstar} gives a histogram of the distance (number of edits)
between two successive war periods. We obtain a mean value of $n^*=1300
\pm 90$.

{\bf c) Never-ending wars.} In the evolution of the articles in
this category no permanent, or even temporary, consensus gets ever
built. Articles describing intrinsically highly controversial/hot topics
tend to belong in this category.

We sorted all articles with $M_{\infty}>1000$ in one of the categories (a-c)
and calculated the relative share of each category at a given $M_{\infty}$.
The results are shown in Figure~\ref{chart}. Keeping in mind that less than 1\%
of WP pages is controversial (some 12 k out of 3.2 M in the original data set
have $M \geq 10^3$), we see that only a small fraction of these fit the
`multiple consensuses' category (b), with the majority fitting rather clearly
in the two polarly opposed classes (a) and (c). Quite as expected, with the
growth of $M$ category (a) dies out, since consensus is reached, and only
articles in the never-ending war category remain. While in earlier research we
set the controversiality threshold at $M > 10^3$, Figure~\ref{chart} is
suggestive that there is hope for consensus by natural process as long as $M <
10^6$, while the remaining subjects are truly `bad apples', and it is a
credit to the WP community that such cases are kept to a minuscule proportion
of less than 100 in the entire set of 3.2 M articles.

\subsection*{Talk pages and conflict resolution}

Talk pages (also known as a discussion pages) in WP are supposed to be pages
where editors can discuss improvements to an article or other Wikipedia page
\cite{wiki-talk}. Each article could have its own talk page in addition to
user talk pages, which host more personal discussions. In
Ref.~\cite{schneider2010}, case studies of talk pages of 58 selected articles
were reported -- the authors concluded that a considerable portion of talk pages
are dedicated to discussions about removed materials and controversial
edits. In the following, we report our preliminary results on how well talk
pages reflect editorial wars and to what extent they help in resolving
disputes.

\subsubsection*{Talk page length}
 
Those familiar only with the English WP may come to the conclusion that the
length of talk pages associated to each article could provide a simple, direct
measure of controversiality, especially as the whole mechanism of talk pages
was invented to channel controversies.  As can be seen from
Figure~\ref{length-PDF}, article length and talk page length are distributed
quite differently, with the log-normal providing a very good fit for article
length (and a reasonable genesis as a multiplicative process with a left
barrier, here the minimum length of an article, \cite{champernowne1973}) but not
for talk page length, which is no surprise, since there is no left barrier for the
lengths of the talk pages.  (We mention
that the total number of edits on an article has also been argued to be
log-normally distributed \cite{wilkinson2007}.)

As can be seen from Figure~\ref{page-talk}, the correlation between article and
talk page length is not very strong ($C=0.26$) -- the most natural hypothesis
is that the discrepancy is caused by the fact that articles of the same length
can nevertheless have different degrees of controversiality. In the English WP
talk page length correlates reasonably well ($C=0.54$) with $M$ (see
Figure~\ref{M-talk}), yet in other WPs, talk pages are used far less: for
example, in the Hungarian WP editors solve their conflicts directly on the
pages, changing and reverting the versions which they do not like, generally
without any talk or arguments, while in the Spanish WP (which has the longest
talk pages after normalization by article length) or the Czech WP, the talk
pages are generally more cooperative.  According to this result, it becomes
evident that the philosophy of ``talk before type''\cite{viegas2007} is not
truly followed in practice.  Depending on culture, talk pages can be reflective of the
conflicts and edit wars, but they do not act as a dampening mechanism.

\subsubsection*{Discussion networks}

We begin by some qualitative observations that emerge from the manual study of
the networks of editorial interactions such as depicted in
Figure\ref{talk-graph}. It appears that the discussions on talk pages are
dominated by continual back and forth between those editors who hardly change
their opinion. In contrast to other social networks, clusters beyond pairs are
rare. Editors joining the discussion at later stages have very little chance
of becoming one of these high-activity editors. Less active editors tend to
address the more active ones rather than each other -- from studying the text
one gets the distinct impression that they do not consider the other low
activity editors worthy of commenting upon. Also, the less prolific editors
appear more negative, more fierce, hysterical in tone, sometimes downright
irrational. Debates rarely conclude on the basis of merit: typically they are
ended by outside intervention, sheer exhaustion, or the evident numerical
dominance of one group.

Based on these observations we hypothesized that most of the editorial war is
carried on by only a few editors.  To check this, we have looked at the {\it
  top 5 ratio}, $r_5(n)$ defined as $M_5(n)/M(n)$ where $M_5$ is the value of
$M$ only considering the contributions of the top 5 pair of editors (ranked by
their mutual reverts) among all the editor pairs of the article.  In
Figure~\ref{M5}, values of $r_5$ calculated from the whole sample are visualized
as a function of $n$ and $M$. The color code is corresponding to the average
value of $r_5$ for the points located in each cell.  Perhaps surprisingly,
this number is quite large ($> 0.5$)) for many articles and for long periods
of the article's life, meaning that a large fraction of the whole war is
indeed caused by a small number of fighting pairs.  $r_5$ becomes smaller than
0.5 only for the articles which are already in the controversial region
($M>1000$) and were edited many times ($n>10^4$).  Smaller values of $r_5$ can
be observed only in the articles which belong to the category of never-ending
wars. In these articles, many different editors have fought at different
periods of time, and a steady flow of replacement armies keeps the article
always far from equilibrium.

In conclusion, we showed that conflicts and editorial wars, although
restricted to a limited number of articles which can be efficiently located,
consume considerable amounts of editorial resources. Moreover, we observed
that conflicts have their own temporal fingerprint which is rooted in memory
effects and the correlation between edits by different editors. Finally, we
demonstrated that, even in the controversial articles, often a consensus can
be achieved in a reasonable time, and that those articles which do not achieve
consensus are driven by an influx of newly arriving editors and external
events. We believe that these empirical results could serve as the basis of
more theoretical agent-centered models which could extend beyond the WP
development process to other large-scale collective and collaborative
problem-solving projects.

\section*{Acknowledgments}

Financial support from EU's~FP7 FET-Open to ICTeCollective
Project~No.~238597 and OTKA grant No.~82333 are acknowledged. We thank Farzaneh~Kaveh and M\'arton~Mesty\'an for helping us perform and validate the
human judgment experiments and Hoda~Sepehri~Rad for useful discussions and technical remarks on implementation of the ranking algorithm.
Comments from an anonymous PLoS~ONE reviewer led to significant
improvements in the presentation and are gratefully acknowledged.   

\bibliography{plosone}

\clearpage
\section*{Figures}

\begin{figure}[!ht]
\begin{center}
\includegraphics[width=0.5\textwidth]{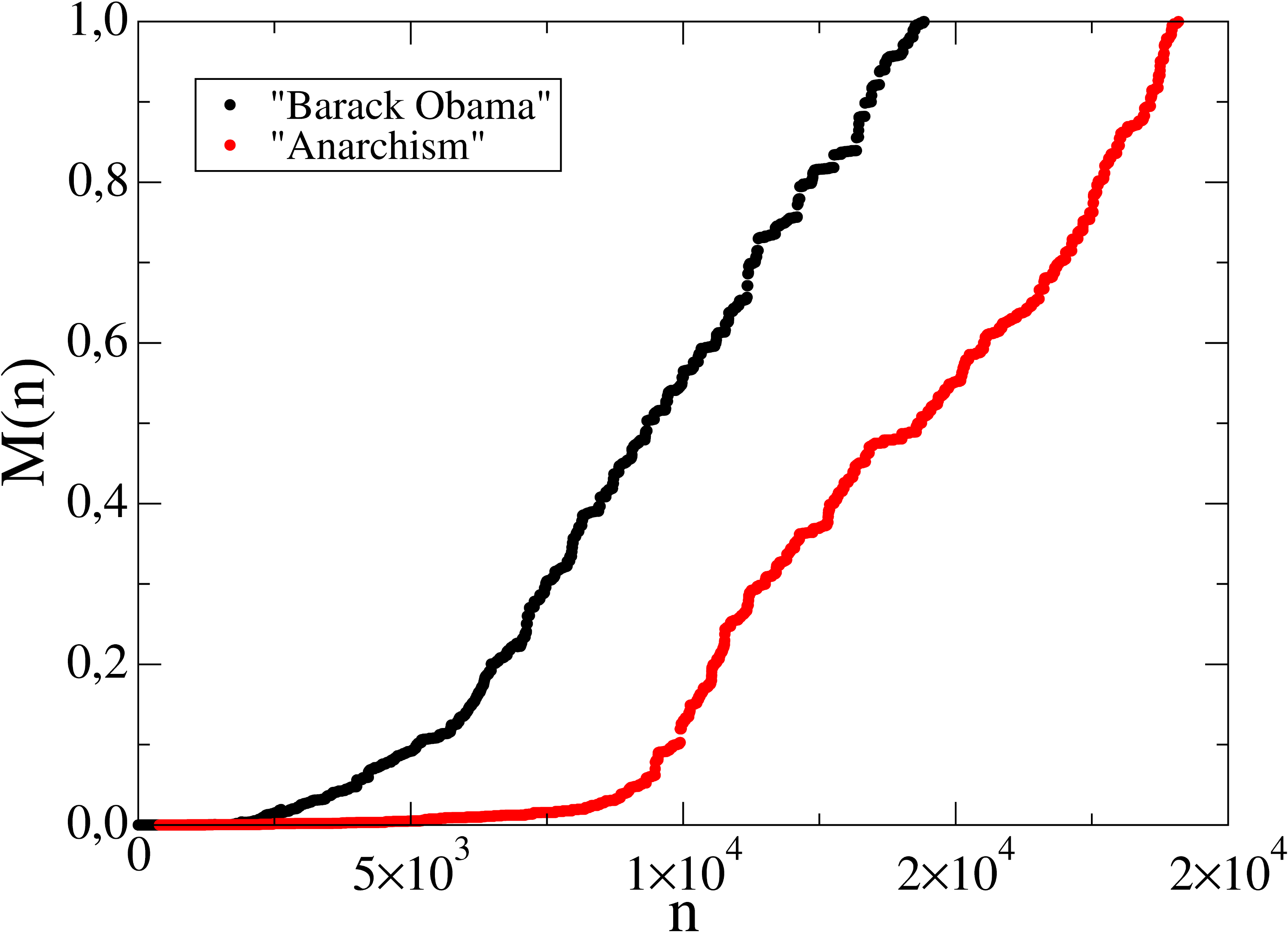}
\caption{{\bf Evolution of controversy measure with number of edits of
    {\tt Anarchism} and {\tt Barack Obama}.} $M(n)$ is normalized to the final value $M_{\infty}$. There is no consensus even for a short period and
editorial wars continue nonstop.}\label{catC}
\end{center}
\end{figure}
\begin{figure}[!ht]
\begin{center}
\includegraphics[width=0.5\textwidth]{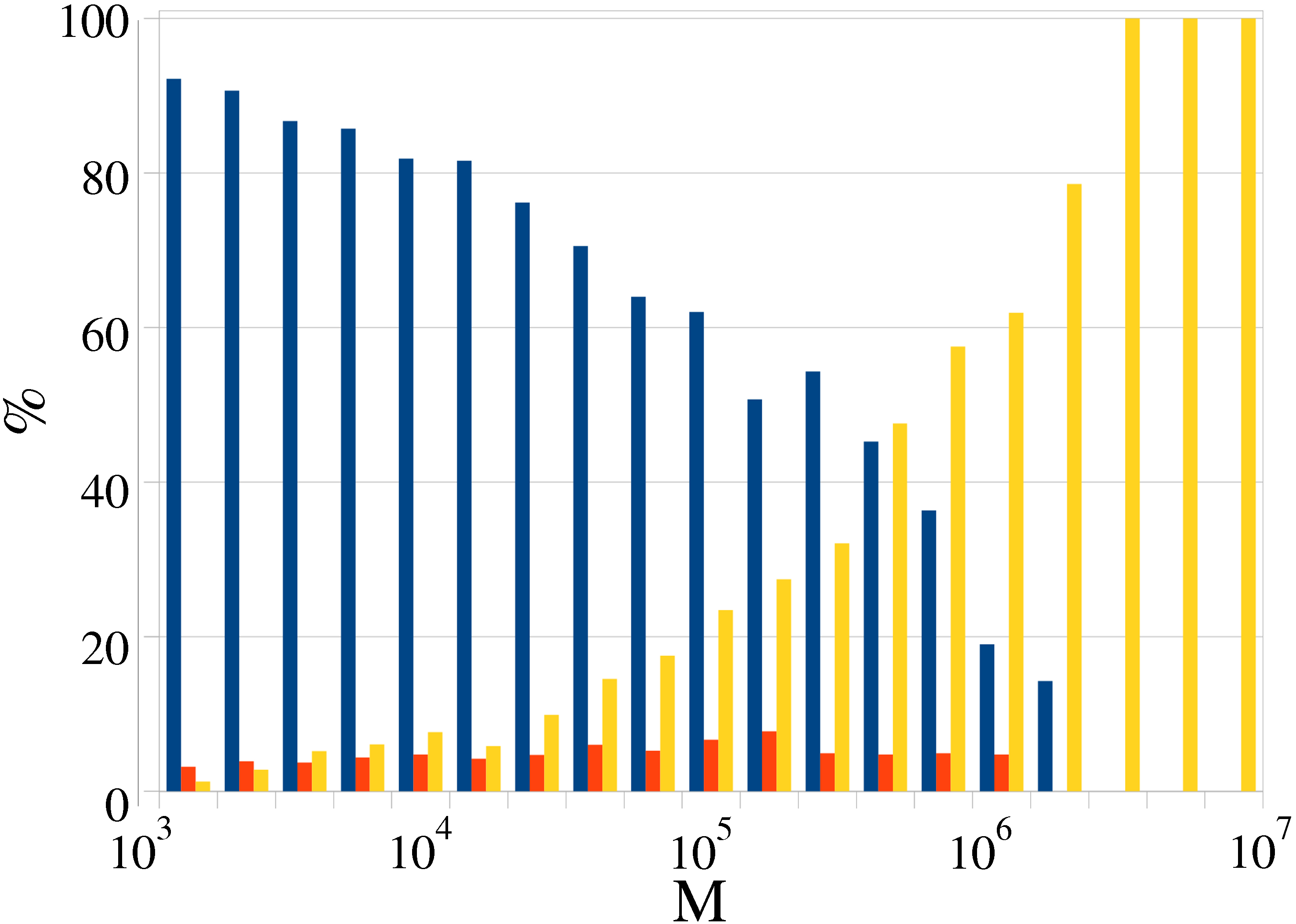}
\caption{{\bf Relative share of each category at different $M$.} Blue: category
(a), consensus. Red: category (b), multi-consensus. Yellow: Category (c), never-ending war.
For the precise definition of each category see the main text.}\label{chart}
\end{center}
\end{figure}
\clearpage
\begin{figure}[!ht]
\begin{center}
 \includegraphics[width=0.75\textwidth]{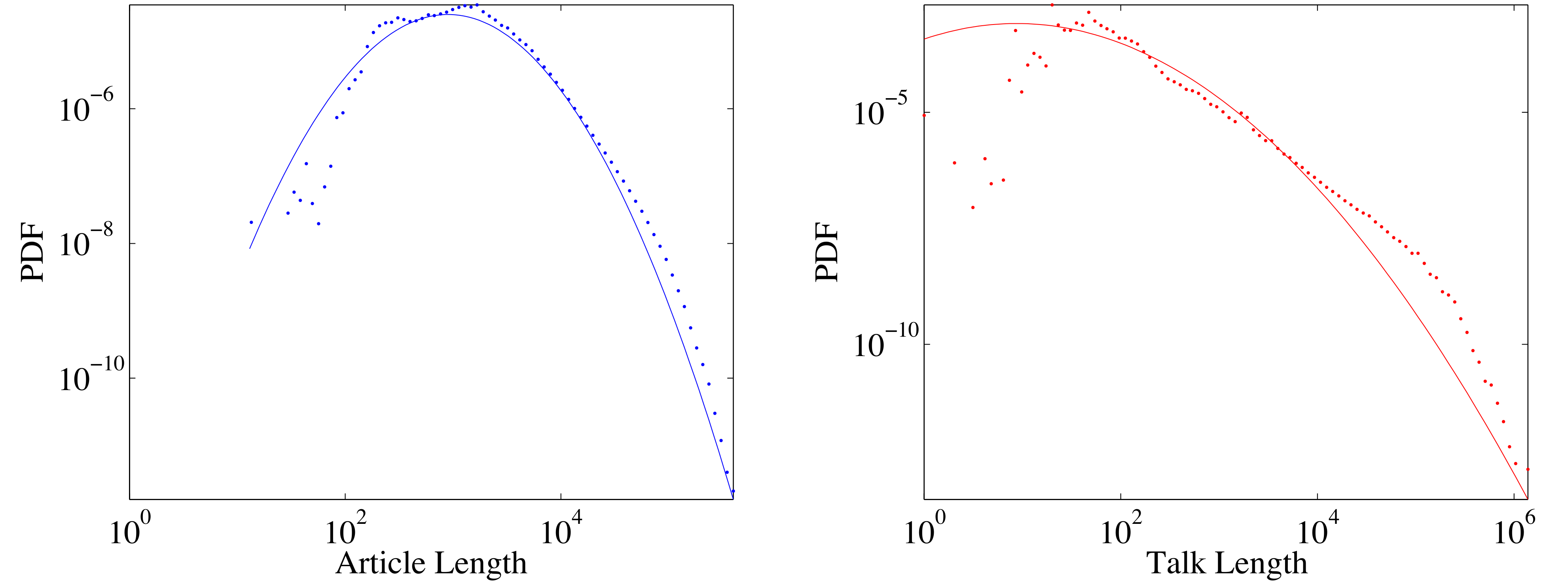}
 \caption{{\bf Length distribution of articles and talk pages with log-normals fits.} The distribution of articles length
is better described by a log-normal distribution compared to the talk length distribution, which tends to be more like
a power-law.}\label{length-PDF}
\end{center}
\end{figure}
\begin{figure}[!ht]
\begin{center}
 \includegraphics[width=0.5\textwidth]{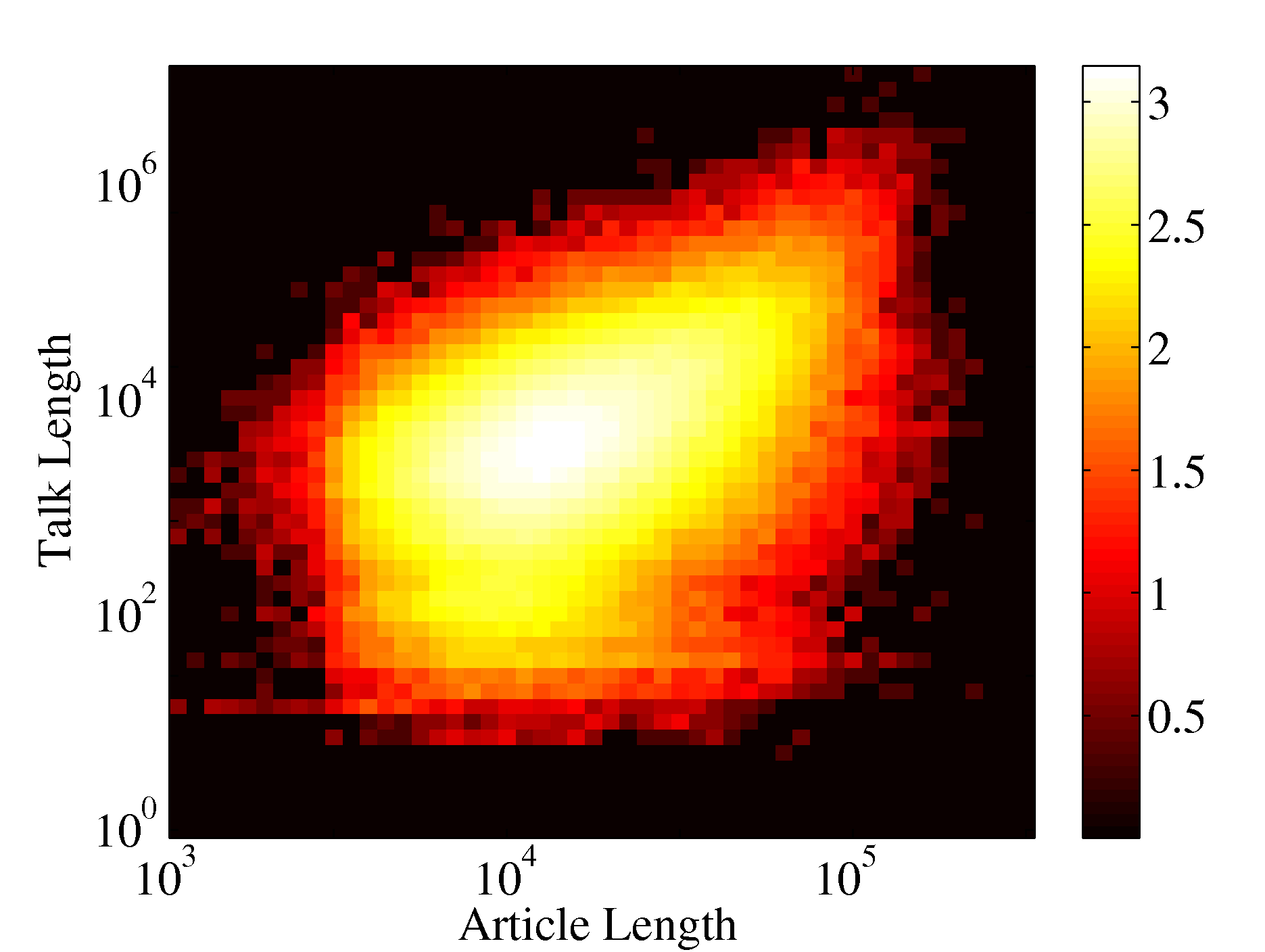}
\caption{{\bf Scatter plot of talk page vs. article length.} Color coding is according to logarithm of the density of points.
The correlation between the length of the article and the corresponding talk
page is weak, $C=0.26$.}\label{page-talk}
\end{center}
\end{figure}
\begin{figure}[!ht]
\begin{center}
 \includegraphics[width=0.5\textwidth]{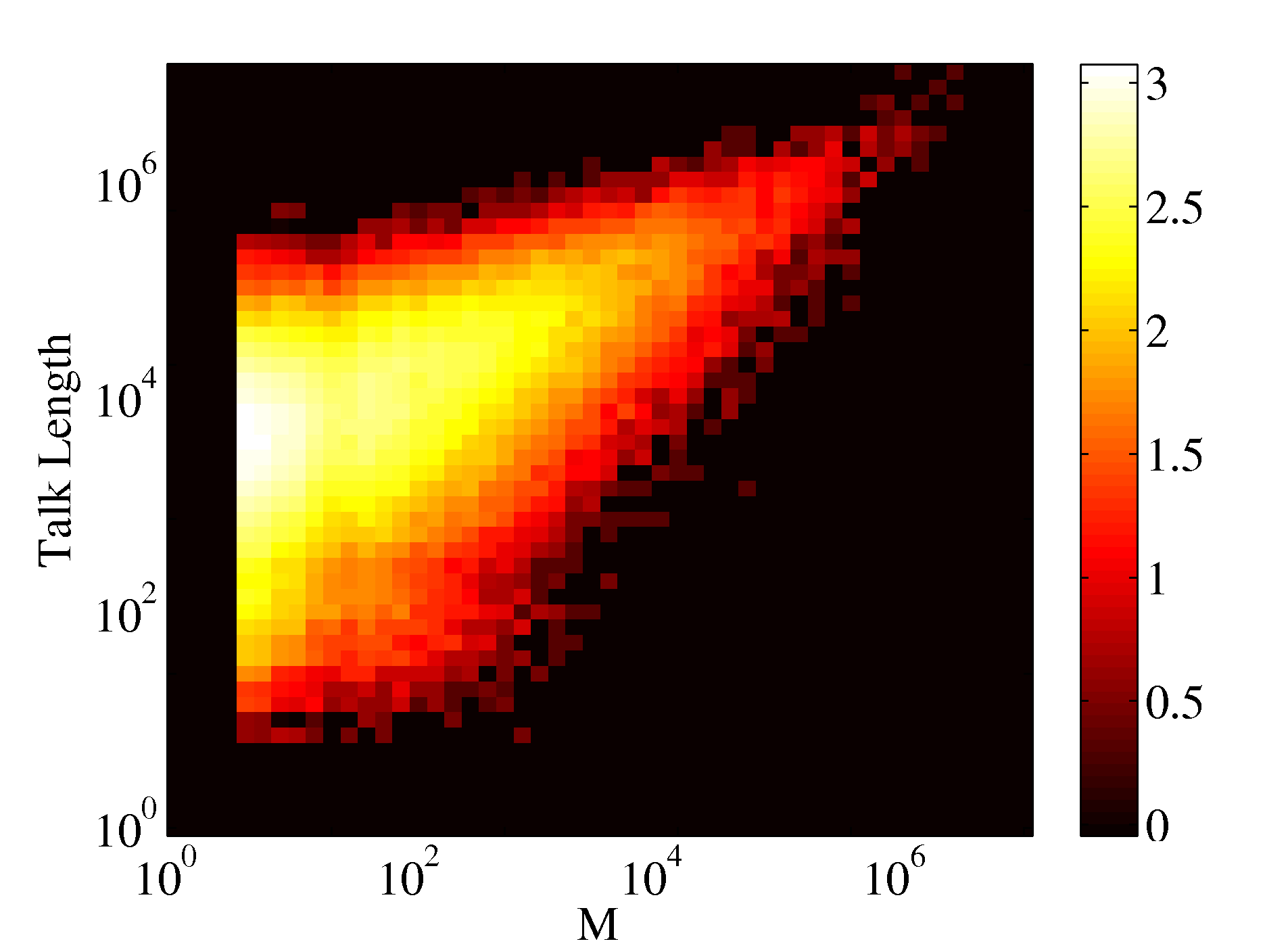}
\caption{{\bf Scatter plot of talk page length vs. $M$.} Color coding is according to logarithm of the density of points.
There is a rather clear correlation, $C=0.54$ between the length of the talk page and the
controversality of the article.}\label{M-talk}
\end{center}
\end{figure}
\begin{figure}[!ht]
\begin{center}
\includegraphics[width=0.50\textwidth]{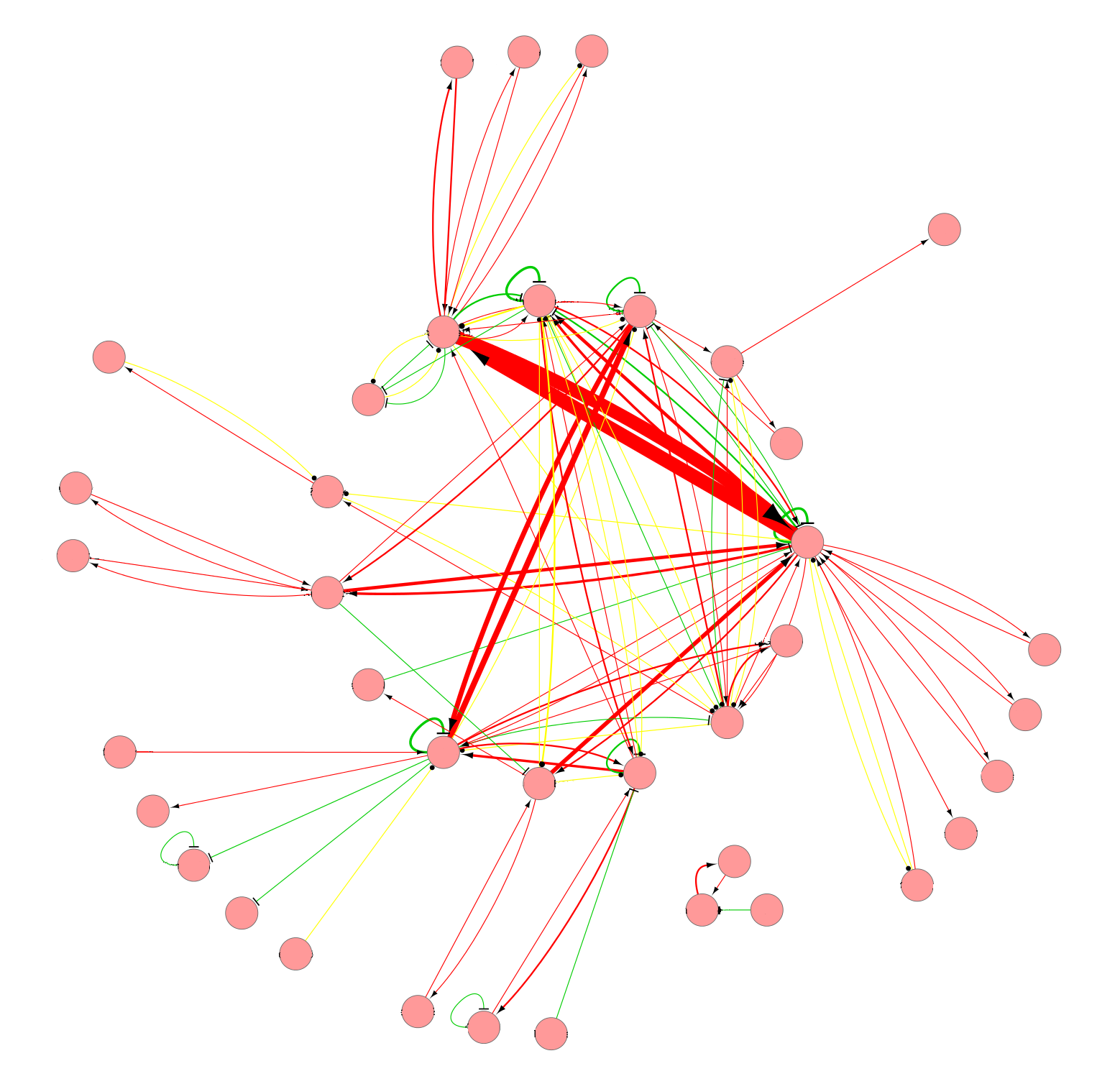}
\caption{{\bf Network representation of editors' interactions in the discussion page of {\tt Safavid dynasty}.}
Each circle is an editor, red arrows represent comments  opposing the target editor,
T-end green lines represent positive comments (agreeing with the other
editor), and yellow lines with round end represent
neutral comments. Line thickness is proportional to the number of times that the same interaction occurs.
Data based entirely on subjective assessments (manual review).} \label{talk-graph}
\end{center}
\end{figure}
\begin{figure}[!ht]
\begin{center}
\includegraphics[width=0.5\textwidth]{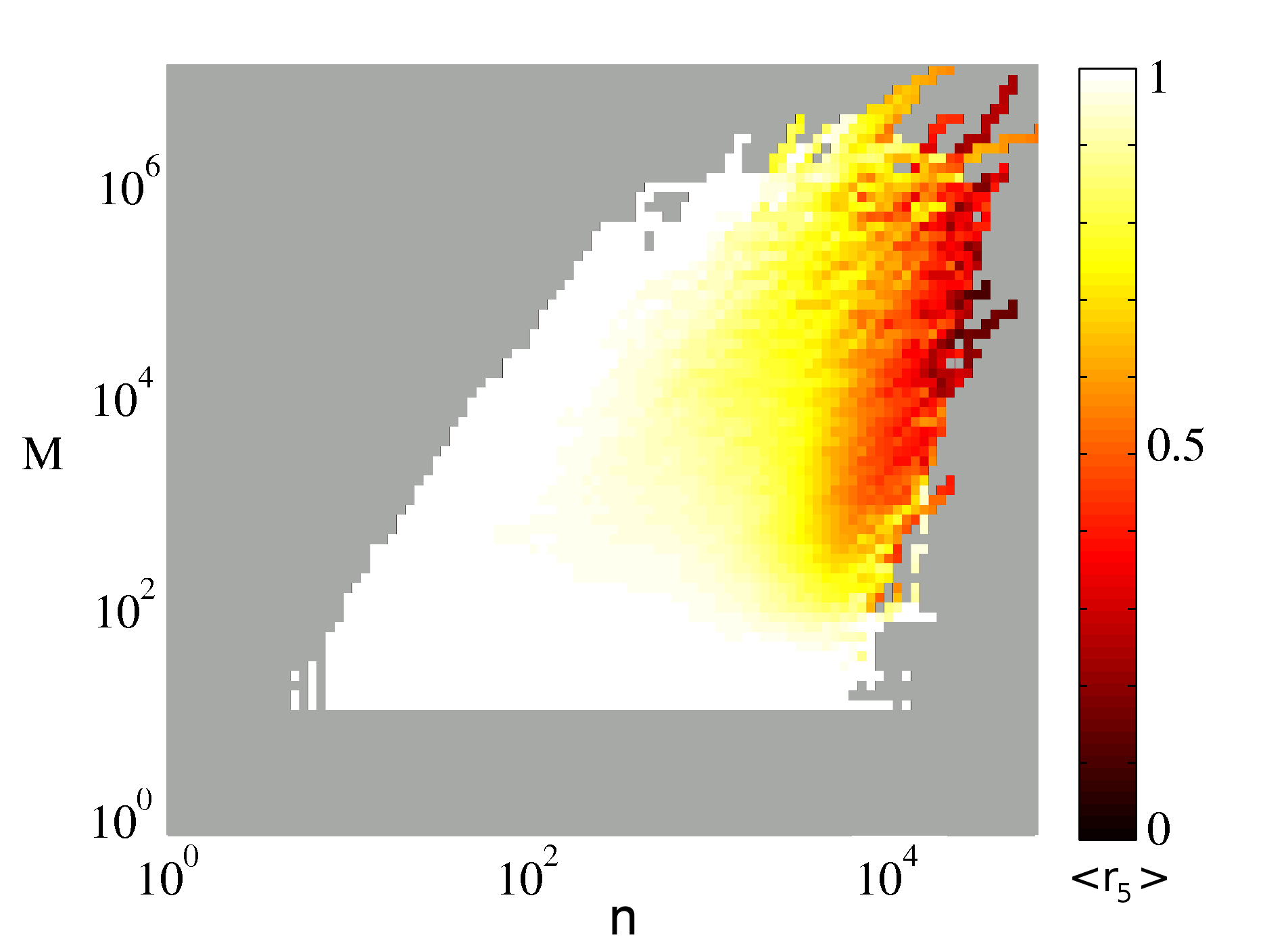}
\caption{{\bf Average $r_5=M_5(n)/M(n)$, color coded for different $M$'s and $n$'s.}  For a
wide range of articles and in a long time of their lives $r_5$, the relative contribution of the top 5 most reverting pair of editors, is very
close to 1, making clear the important role of the top 5 pairs of fighting
editors.}\label{M5}
\end{center}
\end{figure}

\section*{Tables}
\begin{table}[!ht]
\caption{
{\bf Scaling exponents for the two samples of controversial and  peaceful articles, and users.}}
\begin{tabular}{cccc}
 &$\alpha$    &$\beta$    &$\gamma$    \\
 Low M articles    &$0.89\pm0.02$         &-        &-        \\

 High M articles    &$0.56\pm0.01$        &$2.83\pm0.06$        &$0.97\pm0.01$        \\
 Users        &$0.46\pm0.01$        &$3.05\pm0.03$        &$1.44\pm0.01$        \\
\end{tabular}
{ \flushleft
 Edit patterns of controversial
articles and activity patterns of users show all the expected features of bursty correlated processes.}
\label{tab:exponents}
 \end{table}

\section*{Supporting Information} 
\setcounter{figure}{0}            
\makeatletter
\renewcommand{\thefigure}{S\@arabic\c@figure}

\subsection*{Text S1: Details of classification experiments}

During our work we frequently solicited human judgment on how peaceful or
controversial certain pages appear to the observer.  Rather than relying on the
everyday meaning of {\it peaceful} versus {\it controversial}, we have instructed
the judges to use several confluent criteria that we list here in no
particular order.

\begin{itemize}
\item {\bf Rant:} truly hysterical behaviour without much content, usually
  against someone or a group of editors

\item {\bf Help:}  asking for outside help or the help of other editors

\item {\bf Vote:}  voting, merging, moving or talking about these

\item {\bf Prot:} talking about protecting the page

\item {\bf Ban:} talking about banning somebody

\item {\bf Warn:} warning about some bad consequences if somebody does
  something

\item {\bf Command:} ordering, rather than asking, somebody to do, and
  especially to not do, something

\item {\bf Rev:} talking about reverts

\item {\bf Irony:} ironizing over the others. This could be very rude when it
  is observed jointly with other symptoms like accusation but could be quite
  sophisticated used by senior editors who are generally very
  neutral (not accusing, warning etc.). Same tag used for any form of
  malicious joking at the expense of others

\item {\bf Acc:} accusing somebody in the talk of POV, not reading comments,
  not understanding them, repeating the same arguments etc.

\item {\bf Rep:} talk about repeating the same problems or arguments
  over and over again

\item {\bf Comp:} complaining about anything, generally about the others'
  behavior

\item {\bf Emo:} using emotion related words in argumentation: e.g. {\it I
  strongly disagree, are you kidding?} etc.

\item {\bf Formal:} using formal naming style: e.g. referring to other user as
  {\it User Tabib} or {\it Mr. Tabib} rather than the usual {\it Tabib}

\item {\bf UTCite:} citing user talk pages

\item {\bf SelfSupp:} writing something than adding some new comments
  immediately

\item {\bf Stepwise:} answering former comment line by line
\end{itemize}

Needless to say, judging many of these criteria is also a highly subjective
matter: who is to say whether a certain passage is ironic, whether it truly
constitutes a warning, or whether it is a rant?  Nevertheless, human judges
showed quite significant correlation with one another (and with the
machine-generated $M$ score, as seen e.g. in Fig~2). In the body of the paper
we reported on experiments that took the high-conflict sample from the range
$10,000 < M < 70,000$ and the low-conflict control from the range $100 < M <
150$, i.e. on the average a factor of 280 between the two groups.

To test how well humans do, we constructed a less sharply separated sample of
30 pages with $M \approx 50$ for low conflict and 30 pages with $M \approx
2,500$ for high conflict, i.e. on the average a factor of 50 between the two
groups. We had four human judges, instructed in the above criteria, who had to
check all 60 pages given to them in random order. The most peace-leaning judge
found 33 instances of controversality, the most war-leaning judge found 39 (in accordance
with our design of the measure $M$, which aimed at generating fewer false
positives than false negatives). Remarkably, the correlation between the most
lenient and the most strict judge is still $r=0.92$, with a $\kappa$
coefficient of $0.79$, at the high end of what is generally considered
`substantial' agreement. (This is the worst case: the correlation between the
most war-leaning judge and the other two judges is $0.935$ and $0.987$,
Cohen's $\kappa$ is $0.82$ and $0.96$, usually considered `almost perfect'
agreement.)

Not only are the opinions of the judges correlated, they show evident
graduality: if the most peace-leaning judge declares a page controversial, the
other three will declare it controversial with probability $1, 0.9,$ and $1$
respectively, and if the most war-leaning judge declares it peaceful the others
will also do so with probability $1, 1,$ and $0.9$. The result is a manifestly
bimodal distribution, where if we assign one point for each vote of
controversality, 49 pages receive 0 or 4 points, another 10 receive 1 or 3
points, and only 1 page in the entire sample of 60 receives 2 points, truly
splitting the judges. Were the judgments uncorrelated, we would expect to see
the exact opposite picture, with most pages (22.5 out of 60) receiving a score
of 2, 15-15 receiving 1 and 3, and less than 8 receiving some extreme score.

Based on this level of interobserver agreement there cannot be any doubt that
manual classification of WP pages as peaceful vs. controversial can be done quite
reliably. This is not to say that the process is completely repeatable, but
even a lot simpler classification tasks, such as deciding weather a character
is an {\it l} or a {\it 1}, or whether a word in some context is a noun or a
verb, tend to fall shy of $r$ or $\kappa > 0.95$. However, we relied on human
judgment only to the extent it was necessary to create, calibrate, and
validate our controversy measure $M$, all subsequent results use $M$ directly
and are therefore fully replicable.

It is perhaps worth pointing out that our primary interest is not with the
human concept of controversality, but rather with the wars themselves.
Accordingly, we have not made an all out effort to minimize the
misclassification rate of $M$, and there is no doubt that by including more
factors (ranging from talk page length to the number of times banning somebody
is discussed) a much more sensitive measure could be developed. However, as
$M$ correlates nearly as well with human judgment as the least correlated
humans correlate with one another, $r=0.80$ vs. $r=0.85$, there is no reason
to believe that a more sensitive measure would substantially alter the picture
presented here.

\begin{figure}[!ht]
\begin{center}
\includegraphics[width=0.45\textwidth]{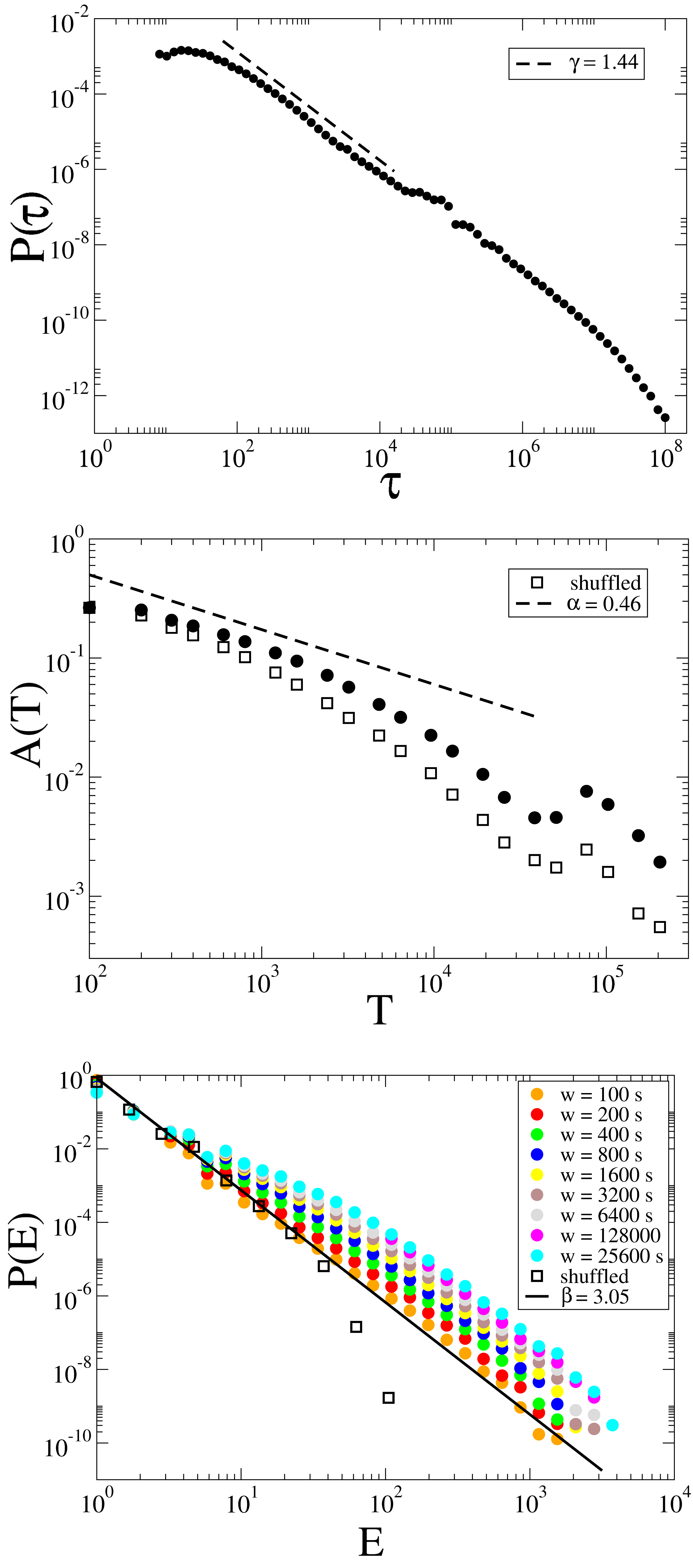}
\caption{{\bf Burst statistics for users' editorial activity}. Upper panel: distribution of
  time interval between two successive edits made by a certain user on any
  article $\tau$. Middle panel: $E$, the number of events in the bursty
  periods separated by a silence window of $w$. Lower panel:
  autocorrelation function $A(T)$ for the editing time train of individual
  users.}\label{user-burst}
\end{center}
\end{figure}
\makeatother

\end{document}